\documentclass[11pt,nofootinbib,preprintnumbers,superscriptaddress]{revtex4}

\usepackage{amssymb, amsfonts, amsmath, bm}
\usepackage{color}
\usepackage{ulem}

\usepackage{mathrsfs}
\usepackage{enumerate}
\usepackage{amsfonts}
\usepackage[
]{graphicx}

\usepackage[
]{hyperref}
\hypersetup{%
 setpagesize=false,
 bookmarksnumbered=true,%
 bookmarksopen=true,%
 colorlinks=true,%
 linkcolor=blue,%
 citecolor=red}

\baselineskip=\normalbaselineskip
\renewcommand{\baselinestretch}{1.4}
\begin{document}
\setcounter{page}{0}

\vspace{25mm}
\title{
{\Large{\bf
Revisiting timelike geodesics \\ in the Fisher/Janis--Newman--Winicour--Wyman spacetime}}
\vspace{10mm}
}
\author{\large \sc Keisuke Ota}
\email{m193405w@st.u-gakugei.ac.jp}
\affiliation{Department of Physics, Tokyo Gakugei University, 4-1-1 Koganei, Tokyo 184-8501, Japan} \affiliation{Tokyo Metropolitan Fukasawa High School, 7-3-14 Fukasawa, Setagaya, Tokyo 158-008, Japan}
\author{\large \sc Shinpei Kobayashi}
\email{shimpei@u-gakugei.ac.jp}
\affiliation{Department of Physics, Tokyo Gakugei University, 4-1-1 Koganei, Tokyo 184-8501, Japan}
\author{\large \sc Keisuke Nakashi}
\email{knakashi@kochi-ct.ac.jp}
\affiliation{Department of Social Design Engineering,
National Institute of Technology, Kochi College, 200-1 Monobe Otsu, Nankoku, Kochi, 783-8508, Japan}
\affiliation{Department of Physics, Rikkyo University, Toshima, Tokyo 171-8501, Japan}
\date{\today}
\preprint{RUP-21-24}
\begin{abstract}
\vspace{10mm}
We investigate the timelike geodesics and the periapsis precession of orbits in the Fisher-Janis-Newman-Winicour-Wyman spacetime.
This spacetime represents the naked singularity spacetime in the Einstein-massless scalar system.
We revisit the results in the previous studies and relax the assumptions about the eccentricity of a bound orbit and the size of a semilatus.
We find that the negative periapsis precession occurs when the spacetime sufficiently deviates from the Schwarzschild spacetime.
In particular, for the small eccentric orbits, we show the negative periapsis precession occurs for $\gamma < 1/2$, where $\gamma$ is the deviation parameter from the Schwarzschild spacetime.
We also obtain the analytical solutions for the special cases of $\gamma=0,1/2,1/4$.
Then, we show that the negative precession never occurs for $\gamma=1/2$.
\end{abstract}

\maketitle

\renewcommand{\baselinestretch}{1.4}

\renewcommand{\thefootnote}{\arabic{footnote}}
\setcounter{footnote}{0}
\addtocounter{page}{1}
\newpage

\section{Introduction}

According to general relativity, at the late stage of the gravitational collapse, the spacetime singularity is generally formed.
It is called the singularity theorem and was proved by Penrose and Hawking~\cite{PhysRevLett.14.57,Hawking:1967ju,Hawking:1970zqf}.
The spacetime singularity that can be observed is called a naked singularity.
If naked singularities are formed frequently, we can distinguish them from black holes by observing characteristic phenomena of the naked singularity.
With regards to this, Penrose proposed the cosmic censorship conjecture~\cite{Penrose:1969pc,1979grec.conf..581P}.
This conjecture states that all spacetime singularities are hidden by black hole horizons:
there cannot be naked singularities.
However, despite many researchers attempt to prove it, no one has mathematically and rigorously proved it yet.
Then, researchers have searched for a candidate of the counterexamples.
In the last few decades, the possibilities of forming naked singularities at the final state of the gravitational collapse in some models have been pointed out (see for example,~\cite{Harada:2001nj,Joshi:2011rlc} and references therein).

There are many spacetimes representing naked singularities as solutions to the Einstein equation.
In this paper, we focus on a specific naked singularity spacetime:
the Fisher-Janis-Newman-Winicour-Wyman (FJNWW) spacetime.
This spacetime was first discovered by Fisher~\cite{Fisher:1948yn},
and it was rediscovered by Janis, Newman, and Winicour~\cite{Janis:1968zz},
and independently by Wyman~\cite{Wyman:1981bd}.
Virbhadra showed Weyman's solution is equivalent to the solution of Janis {\it et al.}~\cite{Virbhadra:1997ie}.
For these reasons,
we call this spacetime the FJNWW spacetime, though it is often called the JNW spacetime.
The FJNWW spacetime is the most general static, spherically symmetric, and asymptotically flat spacetime in the Einstein-massless scalar systems in arbitrary dimensions~\cite{Roberts:1993re,Xanthopoulos:1989kb}.
The FJNWW spacetime reproduces the Schwarzschild spacetime by adjusting the scalar ``charge" to zero, which is a conserved quantity associated with the scalar field.
Therefore, the FJNWW spacetime is useful for discussing the characteristic phenomena of naked singularities.

Since a naked singularity is not covered by a horizon,
it shows some characteristic observational signatures different from the black holes.
For this reason, many researchers have investigated observational signatures
of the FJNWW spacetime, e.g., gravitational lensing~\cite{Virbhadra:1998dy},
the observable images of shadows and thin accretion disks~\cite{Gyulchev:2019tvk,Sau:2020xau},
and the behavior of circular geodesics and the properties of accretion disks~\cite{Chowdhury:2011aa}.
In particular, recent studies have shown that a naked singularity can exhibit a negative periapsis precession~\cite{Dey:2019fpv,Bambhaniya:2019pbr,Joshi:2019rdo,Dey:2020haf,Solanki:2021mkt}.
The negative periapsis precession is that the periapsis precessions in the direction opposite to the direction of a particle motion.
If we find the negative periapsis precession near the gravitational source in future observations, it would be strong evidence of the existence of a naked singularity.
For example, the stellar motions in the Milky Way Galactic Center are continuously observed by GRAVITY and SINFONI~\cite{Abuter:2017,GRAVITY:2020gka,Abuter:2019}.
If there is a naked singularity at the Galactic Center, the periapsis precession may be observed as a smaller value.

In~\cite{Bambhaniya:2019pbr,Joshi:2019rdo,Dey:2020haf}, the authors have investigated the periapsis precession of orbits with small eccentricity in the weak field approximation in the FJNWW spacetime.
Furthermore, in~\cite{Solanki:2021mkt}, the authors numerically calculated and showed that the periapsis precession can take a negative value in the rotating FJNWW spacetime.
In this paper, we revisit the results in~\cite{Joshi:2019rdo}, where the authors focus on orbits with large semilatus and small eccentricity.
We relax the assumptions about the eccentricity or the semilatus of orbits to determine the parameter region in which the negative periapsis precession occurs.
As the result, we show that when the deviation from the Schwarzschild spacetime is large enough,
the negative periapsis precession can occur even if the orbit is far away from the singularity.
Also, for the orbits with the small eccentricity, we show that the negative periapsis precession appears for $\gamma<1/2$, where $\gamma$ is the deviation parameter from the Schwarzschild spacetime, which takes $0< \gamma \le 1$.
As we give the definition in Sec.~\ref{revjnw},
$\gamma$ is related to the ADM mass and the scalar charge,
and the case of $\gamma=1$ corresponds to the Schwarzschild spacetime.

We also discuss analytical solutions to the geodesic equation for the FJNWW spacetime with some special values of the deviation parameter $\gamma$.
For the orbits without any assumptions, numerical calculations are often used when ones examine whether the negative periapsis precession occurs or not (for example, Refs.~\cite{Solanki:2021mkt,Bambhaniya:2020zno}).
Although the numerical analysis is a powerful tool, we can more directly determine the parameter region in which the negative periapsis precession occurs if we can obtain analytical solutions.
In particular, we obtain the analytical solution of the geodesic equation for the case of $\gamma=1/2$ and show that the negative precession never occurs.
The value $\gamma=1/2$ is the critical value of disappearing the negative periapsis precession for the small eccentricity orbits.
This result supports our results that for the orbits with small eccentricity, the negative periapsis precession does not happen for $\gamma \ge 1/2$.

This paper is organized as follows.
In Sec.~\ref{revjnw},
we give a short review of the FJNWW spacetime
and the behavior of geodesics there.
In Sec.~\ref{uande},
we revisit the analysis performed in~\cite{Joshi:2019rdo} and investigate
the conditions in which the periapsis precession becomes negative
with considering the validity of the approximation in detail.
In Sec.~\ref{extension}, to investigate the orbits close to the singularity,
we solve the geodesic equation approximately for either small eccentricity or large semilatus.
In Sec.~\ref{pergamma12}, we derive the exact periapsis precession for the FJNWW spacetime for $\gamma=1/2$ and compare it to each approximate result.
The last section is devoted to the conclusion and the discussion.
In this paper, we work in unit where $c=G=1$.
In the Appendix, we give analytic solutions to the geodesic equations for some particular values of $\gamma$.

\section{A short review of the FJNWW spacetime and timelike geodesics\label{revjnw}
}
In this section, we give a short review of the FJNWW spacetime.
We consider the following action,
\begin{align}
S=\frac{1}{16\pi}\int d^{4} x \sqrt{-g}
\left(R-\frac{1}{2}\partial^\mu\varphi\partial_\mu\varphi\right).
\end{align}
The field equations are written as
\begin{align}
R_{\mu\nu}-\frac{1}{2}Rg_{\mu\nu}=\frac{1}{2}\partial_\mu\varphi\partial_\nu\varphi-\frac{1}{4}g_{\mu\nu}\partial_\alpha\varphi\partial^\alpha\varphi,
\end{align}
\begin{align}
\nabla^\mu\nabla_\mu\varphi=0.
\end{align}
Assuming a static and
spherically symmetric spacetime, we obtain the following metric
of the FJNWW spacetime
\begin{align}
ds^{2}&=-f^{\gamma}d t^{2}+f^{-\gamma} d r^{2}+f^{1-\gamma} r^{2} (d\theta^2+\sin^2\theta d\phi^2),\\
f(r)&=1-\frac{r_g}{r},
\end{align}
where $r_g = 2\sqrt{M^{2} + q^{2}}$, $\gamma = 2M/r_g \in [0,1]$.
$M$ is the ADM mass
and $q$ is  a parameter that relates to the scalar field  $\varphi$ as
\begin{align}
\varphi=\frac{2q}{r_{g}} \log f = \sqrt{1-\gamma^2} \log f.
\end{align}
The  FJNWW spacetime has a naked singularity at $r=r_g$.
Indeed, the Kretschmann invariant and the scalar field diverges at
$r=r_g$ as
\begin{align}
R^{\mu\nu\lambda\sigma}R_{\mu\nu\lambda\sigma}=\frac{r_g^2}{4r^8}f^{2\gamma-4}\bigl[48r^2\gamma^2-16rr_g\gamma(1+\gamma)(1+2\gamma)+r_g^2(1+\gamma)^2(3+2\gamma+7\gamma^2)\bigr],
\end{align}
and it is known that this is a strong, globally naked singularity~\cite{Virbhadra:1995iy}.

Now we review the timelike geodesics in the FJNWW spacetime.
Since the FJNWW spacetime is static and spherically symmetric,
it has the following conserved quantities
\begin{align}
E=f^\gamma \dot{t},\quad
L=r^2 f^{1-\gamma} \sin^2 \theta \dot{\phi},
\label{eq:conservedQs}
\end{align}
where the dot denotes the derivative with respect to the affine parameter.
We normalize the four-velocity $v^{\mu}=\dot{x}^{\mu}$ as $v^{\mu}v_{\mu}=-\epsilon$, where $\epsilon=1$ for timelike geodesics and $\epsilon=0$ for null geodesics, respectively.
The conserved quantities $E$ and $L$ can be interpreted as the energy and the angular momentum per unit rest mass of a test particle, respectively.
Due to the spherical symmetry, we can restrict the particle motions on the equatorial plane ($\theta=\pi/2$) without losing generality.

From the normalization condition $v^{\mu}v_{\mu}=-\epsilon$ together with
Eq.~\eqref{eq:conservedQs}, we obtain
\begin{align}
\dot{r}^2=E^2-V_{\mathrm{eff}}^2,
\end{align}
where $V_{\mathrm{eff}}^2$ is the effective potential given by
\begin{align}
V_{\mathrm{eff}}^2=f^\gamma\left(\epsilon+f^{\gamma-1}\frac{L^2}{r^2}\right).
\end{align}

\begin{figure}[t]
 \centering
 \includegraphics[keepaspectratio,width=170mm]{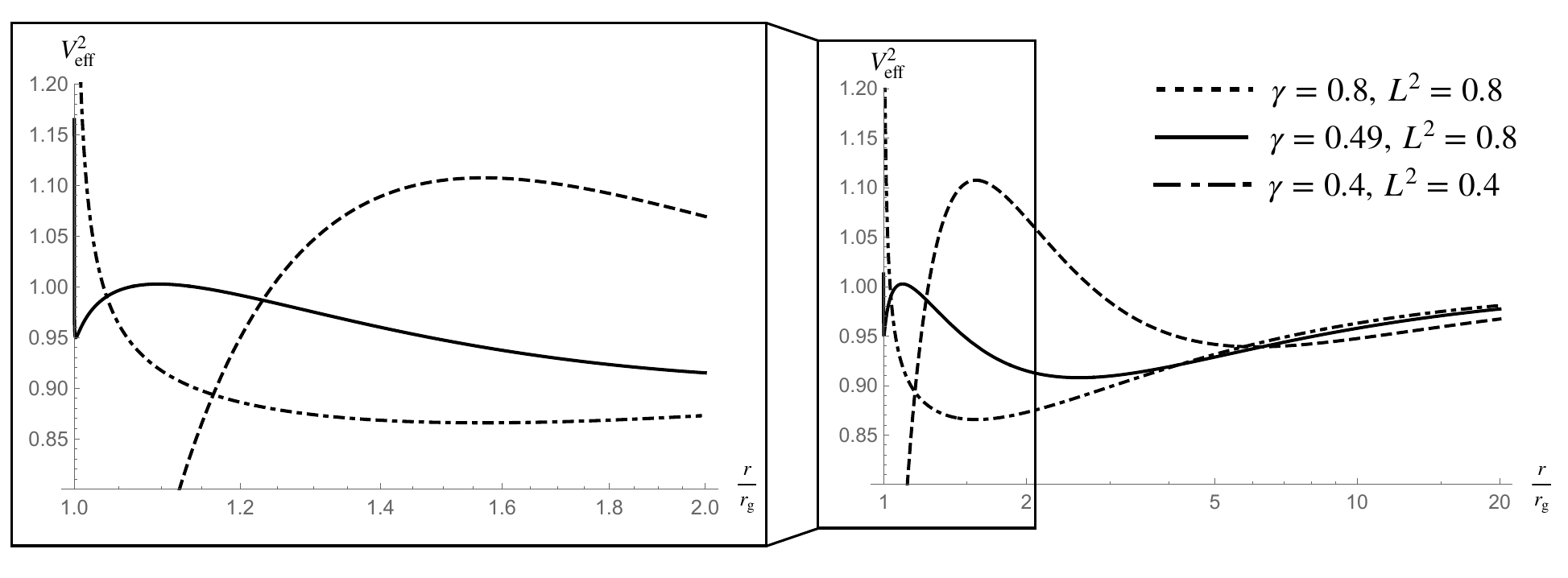}
 \caption{Typical behaviors of the effective potential as a function of radius
 for various values of $\gamma$  and $L$.}
 \label{effectivepotentialbehavior}
\end{figure}

The equation of motion for a test particle
in the FJNWW spacetime
in terms of the derivative with respect to $\phi$ can be written as
\begin{align}
\label{geodesiceq1}
\Bigl(\frac{dr}{d\phi}\Bigr)^2=\frac{r^4}{L^2}f^{2(1-\gamma)}\Bigl(E^2-V_{\mathrm{eff}}^2\Bigr).
\end{align}
In what follows we set $\epsilon=1$
to focus on timelike geodesics.\footnote{
We discuss
null geodesics for particular value of $\gamma$
in the FJNWW spacetime in the Appendix.
}
The typical behaviors of the effective potential $V_{\mathrm{eff}}^2$
for $\epsilon=1$ as a function of radius for various values of $\gamma$  and $L$ are given in Fig.~\ref{effectivepotentialbehavior}.

First we consider a circular orbit as a simple example of geodesics for later use.
The radius of a circular orbit $r_{\mathrm{c}}$ is determined by
the following two conditions:
$V_{\mathrm{eff}}^2(r_\mathrm{c})=E^2$ and $(V_{\mathrm{eff}}^2)'(r_\mathrm{c})=0$.
We rewrite the energy $E$ and the angular momentum $L$ as the functions of $r_{\mathrm{c}}$
as
\begin{align}
E^2&=f^\gamma(r_\mathrm{c})\frac{2r_\mathrm{c}-(1+\gamma)r_g}{2r_\mathrm{c}-(1+2\gamma)r_g},\\
L^2&=r_\mathrm{c}^2f^{1-\gamma}(r_\mathrm{c})\frac{\gamma r_g}{2r_\mathrm{c}-(1+2\gamma)r_g},
\end{align}
which imply that circular orbits exist if $r_{\mathrm{c}}$ satisfies
\begin{align}
r_\mathrm{c}>\frac{r_g}{2}(1+2\gamma)\equiv r_{\mathrm{ph}},
\end{align}
where $r_\mathrm{ph}$ is the radius of the photon sphere.
$r_\mathrm{ph}$ must be larger than $r_g$  for the photon sphere to exist,
which is satisfied if $\gamma > 1/2$.

In addition, by imposing the stability condition $(V_{\mathrm{eff}}^2)''(r_\mathrm{c})>0$, we obtain another condition for the radius of a stable circular orbit
\begin{align}
\label{stableorbitcondition}
r_{\mathrm{c}}<r_- \quad  \text{or} \quad r_+ < r_{\mathrm{c}},
\end{align}
where
\begin{align}
r_{\pm}\equiv\frac{r_g}{2}(3\gamma+1\pm\sqrt{5\gamma^2-1}).
\end{align}
Clearly $1/\sqrt{5} \leq \gamma \ (\leq 1)$ is necessary
for $r_+$ and $r_-$ to exit.
If $r_-<r_g<r_+$, then $r_{+}$ corresponds to the radius of the innermost stable circular orbit.
On the other hand, if $r_g < r_-$, then there are two marginally stable circular orbits.
The sequence of the stable circular orbits is divided into two parts,
which is not seen in the case of the Schwarzschild spacetime.

\begin{figure}[t]
 \centering
 \includegraphics[keepaspectratio,width=90mm]{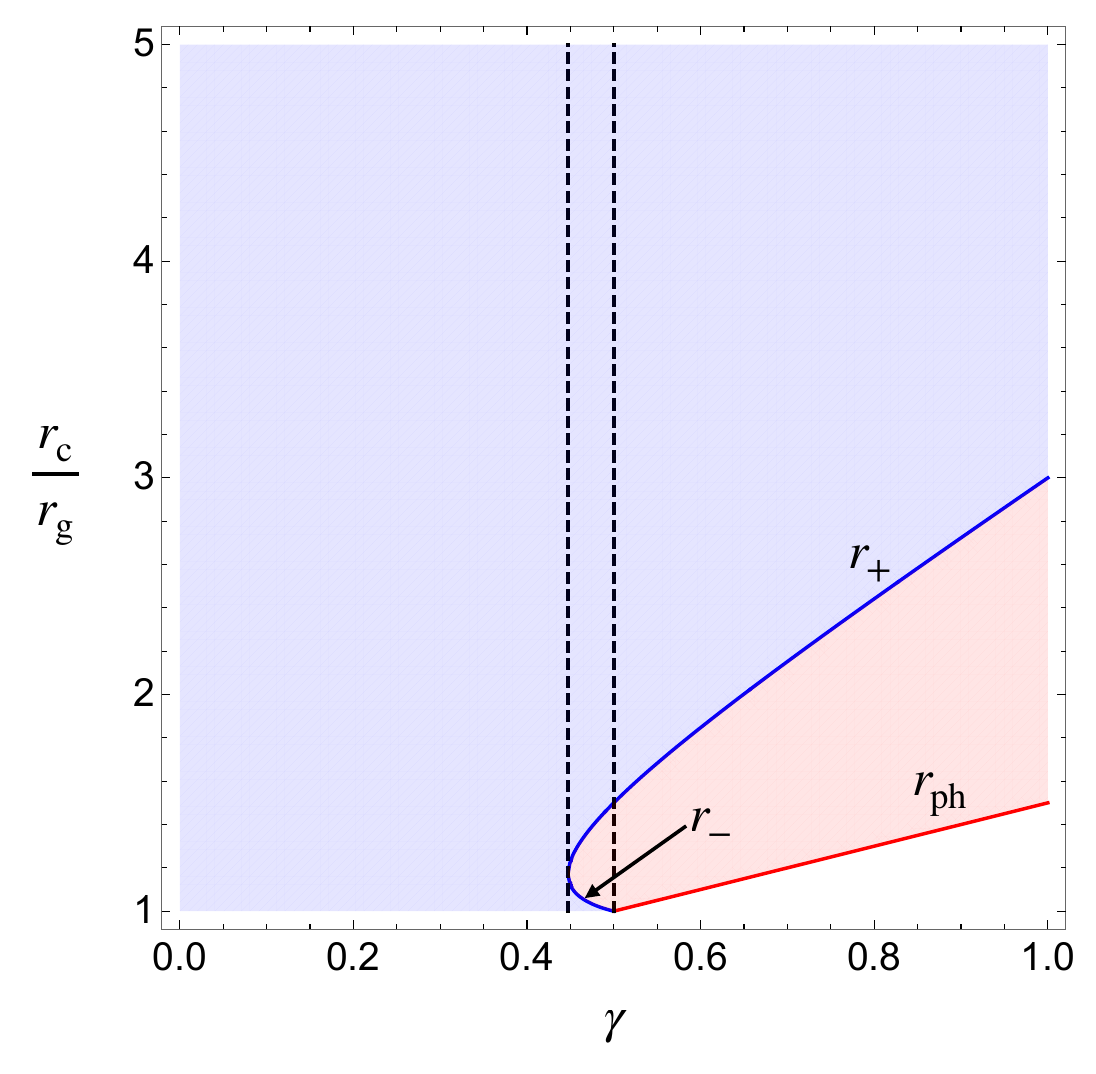}
 \caption{Sequence of stable/unstable circular orbits in the ($\gamma,r_{\mathrm{c}}/r_{\mathrm{g}}$) diagram.
 The red curve and the blue curve correspond to $r_\mathrm{ph}$ and $r_\pm$, respectively.
 The black dashed lines correspond to $\gamma = 1/\sqrt{5}$ and $\gamma=1/2$. The blue region is the sequence of the stable circular orbit.
 In the red region, there are only unstable circular orbits.
}
 \label{geost}
\end{figure}

The geodesics in the FJNWW spacetime have been studied~\cite{Chowdhury:2011aa,Zhou:2014jja}.
They showed that this spacetime has two critical values for $\gamma$. These values are also important in calculating the periapsis precession.
Figure~\ref{geost} shows the radial coordinate values for $r_\mathrm{ph}$, $r_\pm$ as the function of $\gamma$.
We find two critical values $\gamma=1/2,\ 1/\sqrt{5}$, which are the boundaries of the following three regions:

$1.$ $1/2\leq\gamma<1$: \ \ there are one stable circular orbit and one unstable circular orbit\footnote{For $\gamma=1/2$, the effective potential takes a finite value at the singularity.},

$2.$ $1/\sqrt{5}<\gamma<1/2$: \ \ there are two stable circular orbits and one unstable circular orbit,

$3.$ $0<\gamma\leq1/\sqrt{5}$: \ \ there is one stable circular orbit
with arbitrary radius.\\
The three graphs of the effective potential
shown in Fig.~\ref{effectivepotentialbehavior}
correspond to the above three regions.

\section{Periapsis precession of the orbit with small eccentricity in the weak field approximation \label{uande}}
\subsection{Revisiting the previous studies so far}

The definition of the periapsis precession is given by
\begin{align}
\label{periapsisshift}
\Delta_{\text{per}}=2\int_{r_{\text{per}}}^{r_{\text{ap}}}\frac{d\phi}{dr}dr-2\pi,
\end{align}
where $r_{\text{per}}$ and $r_{\text{ap}}$ are the radii of a periapsis and an apoapsis, respectively.
These radii are the real solutions of $E^2-V_{\mathrm{eff}}^2(r)=0$.

The negative periapsis precession ($\Delta_{\text{per}}<0$) occurs when a particle travels smaller than $2\pi$  between two successive periapsis points.
Using the geodesic equation~\eqref{geodesiceq1},
we see that the periapsis precession in the FJNWW spacetime becomes
\begin{align}
\label{jnwperiapsisshift}
\Delta_{\text{per}}=2\int_{r_{\text{per}}}^{r_{\text{ap}}}\frac{L dr}{r^2f^{1-\gamma}\sqrt{E^2-f^\gamma-\frac{L^2}{r^2}f^{2\gamma-1}}}-2\pi.
\end{align}
To show that the negative periapsis precession occurs for appropriate values of $\gamma$,
the authors of~\cite{Joshi:2019rdo} introduced some approximations.
In this section we revisit the analysis performed there and
consider the validity of the approximation.

To this end, we first differentiate Eq.~\eqref{geodesiceq1} with respect to $\phi$
and obtain
\begin{align}
\label{geodesiceq2}
\frac{d^2u}{d\phi^2}=-u+\frac{3}{2}u^2(1-\gamma)\mu\lambda (1-u)^{1-2\gamma}+
\frac{\lambda}{2}(2-\gamma)(1-u)^{1-\gamma},
\end{align}
where $u=r_g/r$, $\mu=E^2$ and $\lambda=r_g^2/L^2$.
Following~\cite{Joshi:2019rdo}, we expand Eq.~\eqref{geodesiceq2} up to the second order of $u$.
This corresponds to the weak field approximation : $r \gg r_g$.
Then we find
\begin{align}
\label{jnwweakeq1}
\frac{d^2u}{d\phi^2}=f_0-f_1u+f_2u^2+\mathcal{O}(u^3),
\end{align}
where
\begin{align}
\label{ff}
f_0&=-\mu\lambda(1-\gamma)+\frac{\lambda}{2}(2-\gamma),\\
f_1&=-\mu\lambda(1-\gamma)(1-2\gamma)+\frac{\lambda}{2}(2-\gamma)(1-\gamma)+1,\\
f_2&=\mu\lambda\gamma(1-\gamma)(1-2\gamma)-\frac{\lambda}{4}\gamma(2-\gamma)(1-\gamma)+\frac{3}{2}.
\end{align}

To focus on the bound orbits with small eccentricities,
we assume a solution for Eq.~\eqref{jnwweakeq1} as
\begin{align}
\label{appsln}
u=u_0[1+e\cos(m\phi)+\mathcal{O}(e^2)],
\end{align}
where $u_0$ is the inverse of the semilatus rectum, and $e~(\ll1)$ is the eccentricity.
Intuitively the inverse of $u_{0}$ can be regarded as the size of the orbit.
Substituting the approximate solution~\eqref{appsln} into Eq.~\eqref{jnwweakeq1} and considering up to the first order of $e$, we can write $u_0$ in terms of $f_0, f_1$ and $f_2$ as
\begin{align}
\label{u01}
u_0=\frac{1}{2f_2}\Bigl(f_1-\sqrt{f_1^2-4f_0f_2}\Bigr).
\end{align}
We also can calculate the value of the periapsis precession as
\begin{align}
\label{perofm}
\Delta_{\text{per}}=2\pi(m^{-1}-1),
\end{align}
where
\begin{align}
\label{m}
  m = (f_{1}^{2}-4 f_{0}f_{2})^{\frac{1}{4}}.
\end{align}
$m$ represents whether the periapsis precession becomes positive or negative.
Actually, if $m$ is larger than unity, the periapsis precession becomes negative.

\subsection{Validity of the approximate solution}

In~\cite{Joshi:2019rdo}, the authors gave the diagram of the parameter region for the negative periapsis precession by changing the scalar charge $q$, the angular momentum $L$ and the energy $E$.
However, in that diagram, there are some geodesics with $u_0 \sim \mathcal{O}(1)$,
which are not valid for the weak field approximation.

To improve this point, we use $e$ and $u_0$ directly
rather than $L$ and $E$
to restrict to the suitable parameter spaces that are physical and
are not in contradiction to the approximation.
Substituting a solution of the  approximated form~\eqref{appsln} into
the geodesic equation up to the order of $u^3$
\begin{align}
\label{jnwweakeq2}
\Bigl(\frac{du}{d\phi}\Bigr)^2=(\mu-1)\lambda+2f_0u-f_1u^2+\frac{2}{3}f_2u^3+\mathcal{O}(u^4),
\end{align}
and evaluating it up to the first order of $e$, we obtain
\begin{align}
\label{u02}
0=(\mu-1)\lambda+2f_0u_0-f_1u_0^2+\frac{2}{3}f_2u_0^3.
\end{align}
Combining Eqs.~\eqref{u01} and~\eqref{u02} for $\mu$ and $\lambda$,
we can write $\mu$ and $\lambda$ in terms of $u_0$ as
\begin{align}
\label{muofu0}
\mu &= 1-\frac{\gamma}{2}u_0+\frac{\gamma}{4}(2\gamma-1)u_0^2+\mathcal{O}(u_0^3, e^2),\\
\label{lambdaofu0}
\lambda &= \frac{2u_0}{\gamma}+\frac{u_0^2}{\gamma}(1-4\gamma)
+\mathcal{O}(u_0^3, e^2).
\end{align}

Substituting Eqs.~\eqref{muofu0} and~\eqref{lambdaofu0} into Eq.~\eqref{m} and expanding up to the second order of $u_0$ (and up to the first order of $e$),
we can see the explicit $u_0$ dependence of $m$ as
\begin{align}
\label{behaviorofm}
m=1-\frac{3}{2}\gamma u_0-\frac{1}{8}(5\gamma^2+6\gamma-2)u_0^2
+\mathcal{O}(u_0^3, e^2).
\end{align}
Then, the periapsis precession as the function of $u_0$ is given by
\begin{align}
\label{periapsisshift1app}
\Delta_{\text{per}}=3\pi\gamma u_0-\frac{\pi}{4}(2-6\gamma-23\gamma^2)u_0^2+\mathcal{O}(u_0^3, e^2).
\end{align}
Clearly the negative periapsis precession occurs if $\gamma$ satisfies
\begin{align}
\gamma<\frac{u_0}{6}\Bigl(1-\frac{u_0}{2}\Bigr)+\mathcal{O}(u_0^3, e^2).
\end{align}
This result shows that
the value of $\gamma$ is sufficiently small, i.e., the scalar charge $q$
is large enough, the negative periapsis precession occurs
even far away from the singularity.

\section{Proofs of the existence of negative periapsis precession
via other approximate methods\label{extension}}

In the previous section, we have evaluated the periapsis precession for the orbits with small eccentricity under the weak field approximation.
In this section,
we determine the maximum value of $\gamma$ where the periapsis precession becomes the negative.
To this end, we allow either arbitrary eccentricity $e$ or arbitrary semilatus $u_0$:
the weak field approximation with arbitrary eccentricity
($u_0 \ll 1$ and $\mbox{}^{\forall} e$)
or the small deviation from a circular orbit with arbitrary semilatus
($e \ll 1$ and $\mbox{}^{\forall} u_0$).

\subsection{The periapsis precession
in the weak field approximation with arbitrary eccentricity \label{onlyu}}

Now we deal with bound orbits with arbitrary eccentricity.
Following~\cite{darwin1959gravity,PhysRevD.50.3816},
 we introduce $u_{\text{per}}$ and $u_{\text{ap}}$ by
\begin{align}
u_{\text{per}}=u_0(1+e),\ \ u_{\text{ap}}=u_0(1-e).
\end{align}
instead of $u_0$ and $e$, where $u_{\text{per}}$ and $u_{\text{ap}}$ are the inverse of dimensionless radius of the periapsis $r_{\text{per}}/r_g$ and the apoapsis $r_{\text{ap}}/r_g$, respectively.
Then, Eq.~\eqref{jnwweakeq2} is factorized to
\begin{align}
\left(\frac{du}{d\phi}\right)^2=\frac{2}{3}f_2(u-u_{\text{ap}})(u-u_{\text{per}})(u-u_1)+\mathcal{O}(u^4).
\end{align}
By comparing the coefficients of the rhs of Eq.~\eqref{jnwweakeq2}
by each order of $u$, we find
\begin{align}
\label{muofu02}
\mu =& \ 1-\frac{\gamma}{2}(1-e^2)u_0+\frac{\gamma}{4}(1-e^2)^2(2\gamma-1)u_0^2+\mathcal{O}(u_0^3), \\
\label{lambdaofu02}
\lambda = &\ \frac{2u_0}{\gamma}+\frac{u_0^2}{\gamma}[1-4\gamma e^2(1-2\gamma)]+\mathcal{O}(u_0^3), \\
\label{u1ofu02}
u_1 = &\ 1+\frac{1}{3}(1-7\gamma^2)u_0 \nonumber \\
&+\frac{1}{18}(1-\gamma)[11+11\gamma-80\gamma^2-98\gamma^3+3e^2(1-2\gamma)(1+3\gamma)]u_0^2+\mathcal{O}(u_0^3).
\end{align}
The periapsis precession is calculated as
\begin{align}
\label{periapsisshiftapp2}
\Delta_{\text{per}}
&=\sqrt{\frac{6}{f_2}}\int_{u_{\text{ap}}}^{u_{\text{per}}}\frac{du}{\sqrt{(u-u_{\text{ap}})(u_{\text{per}}-u)(u_1-u)}}-2\pi \\
&=\sqrt{\frac{6}{f_2}}\int^{\pi}_{0}\frac{d\chi}{\sqrt{u_1-u_0(1+e\cos\chi)}}-2\pi,
\end{align}
where we defined a new variable $\chi$ as
\begin{align}
\label{variabletransf1}
u=u_0(1+e\cos \chi),
\end{align}
and transformed the integral into the familiar form.
Performing another variable transformation $\chi=-2\psi-\pi$, we can write down
$\Delta_{\text{per}}$ using the complete elliptic integral of the first kind $K(\tilde{k})$ \footnote{
In this paper, we use the definition of the complete elliptic integral of the first kind in terms of the parameter $\tilde{k}$ as
\begin{equation}
K(\tilde{k})= \int_0^{\pi/2} \frac{d\psi}{\sqrt{1-\tilde{k} \sin^2 \psi}},
\end{equation}
instead of the definition
\begin{equation}
K(k)= \int_0^{\pi/2} \frac{d\psi}{\sqrt{1-k^2 \sin^2 \psi}},
\end{equation}
where the elliptic modulus $k$ is used.
} as
\begin{align}
\label{periapsisshift21}
\Delta_{\text{per}}=\sqrt{\frac{24}{f_2(u_1-u_0+u_0e)}}
K\left(\frac{2u_0e}{u_1-u_0+u_0e}\right)-2\pi.
\end{align}

The expansion of Eq.~\eqref{periapsisshift21} up to the second order of $u_{0}$
becomes
\begin{align}
\label{periapsisshift21app}
\Delta_{\text{per}}=3\pi\gamma u_0-\frac{\pi}{8}[4-12\gamma-46\gamma^2+(1-12\gamma+8\gamma^2)e^2]u_0^2+\mathcal{O}(u_0^3).
\end{align}
This coincides with the periapsis precession~\eqref{periapsisshift1app} for $e=0$.

If the periapsis precession is negative up to the second order of $u_0$,
then $\gamma$ is bounded from above as
\begin{align}
\gamma<\frac{u_0}{24}(4+e^2)\left(1-\frac{u_0}{2}(1+e^2)\right)+\mathcal{O}(u_0^3).
\end{align}
As the eccentricity approaches unity, the periapsis gets closer to the singularity.
Therefore we conclude that
the negative periapsis precession occurs for larger $\gamma$
in the weak field approximation with arbitrary eccentricity $e$.

\subsection{The periapsis precession of orbits slightly deviated from
a circular orbit with arbitrary semilatus \label{onlye}}

Next, we consider bound orbits slightly deviated from a circular orbit
with arbitrary semilatus.
To this end, we solve the geodesic deviation equation from a circular orbit~\cite{Kerner:2001cw}.
Contrary to the previous subsection, we do not assume
the weak field approximation, i.e., $u_0 \ll 1$.

The geodesic deviation describes the deviation of two adjacent particle orbits.
We represent geodesics with $x^\mu(s,p)$ where $s$ is the affine parameter and $p$ is the label of each geodesic belonging to a smooth geodesic congruence.
The four-velocity and the deviation vector are defined as
\begin{align}
v^\mu=\frac{\partial x^\mu}{\partial s},\quad
n^\mu=\frac{\partial x^\mu}{\partial p},
\end{align}
respectively.
The geodesic deviation can be used to describe the geodesics $x^\mu(s,p)$ close to a given geodesic $x^\mu(s,p_0)$.
By performing the Taylor expansion around $p=p_{0}$, we obtain
\begin{align}
\label{appmotion}
x^\mu(s,p)&=x^\mu(s,p_0)+(p-p_0)\left.\frac{\partial x^\mu}{\partial p}\right|_{s,p=p_0}+\mathcal{O}((p-p_0)^2)\\
&=x^\mu(s,p_0)+(p-p_0)n^\mu(s)+\mathcal{O}((p-p_0)^2).
\end{align}
In order to find the deviation vector $n^{\mu}$, we solve the geodesic deviation equation:
\begin{align}
\label{covdev}
v^\beta \nabla_\beta (v^\alpha\nabla_\alpha n^\mu)=R^\mu_{\ \alpha\beta\gamma}v^\alpha v^\beta n^\gamma.
\end{align}

Here we consider a generic form of a static and spherically symmetric spacetime:
\begin{align}
\label{gnmetric}
ds^2=-A(r)dt^2+A(r)^{-1}dr^2+R^2(r)(d\theta^2+\sin^2\theta d\phi^2).
\end{align}
We solve the geodesic deviation equation~\eqref{covdev}
around a circular orbit of radius $r_\mathrm{c}$.
Then each component of the four-velocity $v^{\mu}$ is given by
\begin{align}
v_\mathrm{c}^t=\frac{E_\mathrm{c}}{A(r_\mathrm{c})},\ v_\mathrm{c}^r=0,\ v_\mathrm{c}^\phi=\frac{L_\mathrm{c}}{R^2(r_\mathrm{c})}\equiv\omega_\mathrm{c},\ v_\mathrm{c}^\theta=0,
\end{align}
where we set $\theta=\pi/2$, and $E_\mathrm{c}$ and $L_\mathrm{c}$
are solutions of $\tilde{V}_{\mathrm{eff}}^2(r_\mathrm{c})=E_\mathrm{c}^2$ and $(\tilde{V}_{\mathrm{eff}}^2)'(r_\mathrm{c})=0$.
They are explicitly given by
\begin{align}
E_\mathrm{c}\equiv\sqrt{\frac{2A^2(r_\mathrm{c})R'(r_\mathrm{c})}{2A(r_\mathrm{c})R'(r_\mathrm{c})-A'(r_\mathrm{c})R(r_\mathrm{c})}},\ \
L_\mathrm{c}\equiv\sqrt{\frac{A'(r_\mathrm{c})R^3(r_\mathrm{c})}{2A(r_\mathrm{c})R'(r_\mathrm{c})-A'(r_\mathrm{c})R(r_\mathrm{c})}},
\end{align}
 where $\tilde{V}^2_{\mathrm{eff}}\equiv A(r)(1+L^2/R^2(r))$.
Hereafter in this subsection,
we will omit the argument
because we evaluate all functions at $r=r_\mathrm{c}$.

From Eq.~\eqref{covdev}, we obtain the equation for $n^\theta$
\begin{align}
\label{geodevfortheta}
\frac{d^2n^\theta}{ds^2}+\omega_\mathrm{c}^2n^\theta=0,
\end{align}
and its solution as
\begin{align}
n^\theta(s)=n_0^\theta\cos(\omega_\mathrm{c} s+\vartheta_0).
\end{align}
This harmonic oscillation can be seen as the degrees of freedom of the coordinate transformation with a new $z$ axis that is slightly inclined for the original one.
Therefore, due to the spherical symmetry, we can choose $n^\theta=0$ without loss of generality.

The geodesic deviation equations for other components $n^t$, $n^r$, $n^\phi$
can be expressed as the following matrix form:
\begin{align}
\label{matrixform}
\left(\begin{array}{ccc}
\frac{d^2}{ds^2} & \frac{A'}{A^2}E_\mathrm{c}\frac{d}{ds} &0\\
A'E_\mathrm{c}\frac{d}{ds} & \frac{d^2}{ds^2}+A\omega_\mathrm{c}^2 \left(\frac{A''}{A'}RR'-R'^2-RR''\right)&-2ARR' \omega_\mathrm{c}\frac{d}{ds}\\
0&\frac{2R'}{R}\omega_\mathrm{c}\frac{d}{ds}&\frac{d^2}{ds^2}
\end{array}\right)
\left(\begin{array}{c}n^t\\n^r\\n^\phi\end{array}\right)=\left(\begin{array}{c}0\\0\\0\end{array}\right).
\end{align}
We assume that
the solutions are proportional to $e^{i\omega s}$ ($\omega\in \mathbb{R}$). Then, we obtain a characteristic equation
\begin{align}
\label{chrcteq}
\omega^4\left[\omega^2-\omega_\mathrm{c}^2F(r_\mathrm{c})\right]=0,
\end{align}
where $F(r_\mathrm{c})$ is the test function to determine the periapsis precession which is explicitly given by
\begin{align}
\label{fdef}
F(r_\mathrm{c}) \equiv
\left. A\left[\left(\frac{A''}{A'}-2\frac{A'}{A}\right)RR'+3R'^2-RR''\right] \right|_{r=r_{\mathrm{c}}}.
\end{align}
A nontrivial solution of Eq.~\eqref{chrcteq} is given by
\begin{align}
\label{omega}
\omega=\omega_\mathrm{c}F^{\frac{1}{2}}(r_\mathrm{c}),
\end{align}
which corresponds to the characteristic frequency of a perturbed orbit.

Since the each component of the matrix equation~\eqref{matrixform}
are the second order differential equations, general solutions may contain
\begin{align}
\left(\begin{array}{c}n^t\\n^r\\n^\phi\end{array}\right)=\left(\begin{array}{c}\Delta v^ts+\Delta t\\\Delta v^rs+\Delta r\\\Delta v^\phi s+\Delta \phi\end{array}\right),
\end{align}
where $\Delta x^\mu$ and $\Delta v^{\mu}$ are constant.
Substituting $n^t, n^r$ and $n^\phi$ into Eq.~\eqref{matrixform}, we obtain
three equations for $\Delta v^t, \Delta v^r$ and $\Delta v^\phi$ as
\begin{align}
\label{trivialequationfort}
&\frac{A'}{A^2}E_\mathrm{c}\Delta v^r=0, \\
\label{trivialequationforr}
& A' E_\mathrm{c}\Delta v^t+ A\omega_\mathrm{c}^2 \left(\frac{A''}{A'}RR'-R'^2-RR''\right)(\Delta v^r s+\Delta r)- 2ARR' \omega_\mathrm{c}\Delta v^\phi=0, \\
\label{trivialequationforphi}
&2\frac{R'}{R}\omega_\mathrm{c} \Delta v^r=0.
\end{align}
We can set $\Delta t = \Delta \phi =0$ without loss of generality
due to the static and spherical symmetry  of the spacetime.
From Eq.~\eqref{trivialequationfort} (or Eq.~\eqref{trivialequationforphi}),
we obtain a trivial solution $\Delta v^r=0$ which corresponds to
the orbits without radial velocity.
On the other hand, from Eq.~\eqref{trivialequationforr}, we find a relation between $\Delta v^t$, $\Delta v^\phi$ and $\Delta r$ as
\begin{align}
\label{constantshiftrelation}
\omega_\mathrm{c}^2 \left(\frac{A''}{A'}RR'-R'^2-RR''\right)\Delta r = 2RR' \omega_\mathrm{c}\Delta v^\phi-\frac{A'}{A}E_\mathrm{c}\Delta v^t,
\end{align}
where $\Delta r$ denotes that the shift of radius of the circular orbit
from $r_\mathrm{c}$ to $r_\mathrm{c}+\Delta r$ \footnote{This is called circular perturbation~\cite{Philipp:2016gyq}. We can set $\Delta r =0$  when $\Delta v^t=\Delta v^\phi=0$.}.
The relation~\eqref{constantshiftrelation} gives the value of the shifts of
the energy and the angular momentum due to the shift of the radius
of the circular orbit.

Adding to the geodesic deviation equations, we can utilize the conservation
of energy of a test particle $v^\mu v_\mu=-1$.
Expanding it up to the first order of $\Delta v^t$, $\Delta v^\phi$ and $\Delta r$,
we obtain
\begin{align}
-A(r_\mathrm{c}+\Delta r)\left(v_\mathrm{c}^t+\Delta v^t\right)^2+R^2(r_\mathrm{c}+\Delta r)\left(v_\mathrm{c}^\phi+\Delta v^\phi\right)^2=-1.
\end{align}
This leads to the following relation
\begin{align}
E_\mathrm{c}\Delta v^t=R^2 \omega_\mathrm{c}\Delta v^\phi.
\end{align}
Combining this equation with the relation~\eqref{constantshiftrelation}, we find
\begin{align}
\label{sieurytnsl}
A \omega_\mathrm{c}^2  \left(2\frac{R'}{R}-\frac{A'}{A}\right)^{-1}\left(\frac{A''}{A'}RR'-R'^2-RR''\right)\Delta r=E_\mathrm{c}\Delta v^t=R^2 \omega_\mathrm{c}\Delta v^\phi.
\end{align}
Setting the affine parameter $s$ to zero at the periapsis,
we obtain a general solution of the deviation equation around the circular orbit as
\begin{align}
\label{geodevsln}
n^t&=\Delta v^ts+\frac{A'E_\mathrm{c}}{A^2\omega}n_0^r\sin(\omega s),\\
n^r&=\Delta r-n_0^r\cos(\omega s),\\
n^\phi&=\Delta v^\phi s+2 \frac{R^{\prime} \omega_{0}}{R \omega} n_0^r\sin(\omega s),
\end{align}
where $n_0^r$ is a positive constant.
Substituting Eq.~\eqref{geodevsln} into \eqref{appmotion}, we find
the small shift from the circular orbit as
\begin{align}
\label{a4eoituvnosomi}
t(s)&=\frac{E_\mathrm{c}}{A} s+\Delta v^t s+\frac{A'E_\mathrm{c}}{A^2\omega}n_0^r\sin(\omega s),\\
r(s)&=r_\mathrm{c}+\Delta r-n_0^r\cos(\omega s),\\
\phi(s)&=\omega_\mathrm{c} s+\Delta v^\phi s+ 2\frac{R^{\prime} \omega_{0}}{R \omega} n_0^r\sin(\omega s),
\end{align}
where we used the fact
that $\Delta r$ and $n_0^r$ are arbitrary constants,
and  included $(p-p_0)$ in their definition.
Now, let us investigate the oscillations around the circular orbits of radius $r_\mathrm{c}$.
For simplicity, we set $\Delta v^t=\Delta v^\phi=0$ and $\Delta r =0 $ as well.
The circular orbit and the perturbed orbit have the angular frequencies, $\omega_\mathrm{c}$ and $\omega$, respectively.
When $\omega$ is larger than $\omega_\mathrm{c}$, the perturbed orbit comes back to the initial periapsis before the circular orbit makes one revolution.
In this case the periapsis precession becomes negative.
To find the parameter region where the periapsis precession is negative,
we use the test function~\eqref{fdef}.
In fact, the proper time between two successive periapsises is $s=2\pi/\omega$. Then, we find the periapsis precession as
\begin{align}
\label{geodevperiapsisshift}
\Delta_{\text{per}}=\phi(s=2\pi/\omega)-2\pi=2\pi\left(\frac{\omega_\mathrm{c}}{\omega}-1\right)=2\pi\left[F^{-\frac{1}{2}}(r_\mathrm{c})-1\right].
\end{align}
This agrees with the results obtained in~\cite{fuchs1990deviation,fuchs1990paralleltransport} when $R(r) = r$.

Now, we discuss the case of the FJNWW spacetime,
where the metric functions $A(r)$ and $R(r)$ are
\begin{align}
A(r)=\left(1-\frac{r_g}{r}\right)^{\gamma},\ R(r)= r \left(1-\frac{r_g}{r}\right)^{\frac{1-\gamma}{2}}.
\end{align}
Then the periapsis precession is given by
\begin{align}
\label{geodevper}
\Delta_{\text{per}}=2\pi\left[F^{-\frac{1}{2}}(r_\mathrm{c}; \gamma)-1\right],
\end{align}
where the test function~\eqref{fdef} is
\begin{align}
\label{testF}
F(r_\mathrm{c}; \gamma)=1-\frac{3r_g}{r_\mathrm{c}}+(1-\gamma)\left(1-\frac{r_g}{r_\mathrm{c}}\right)^{-1}\frac{r_g}{r_\mathrm{c}}\left[3-\left(\gamma+\frac{5}{2}\right)\frac{r_g}{r_\mathrm{c}}\right].
\end{align}
The periapsis precession~\eqref{geodevper} coincides with Eq.~\eqref{periapsisshift1app} for $r_\mathrm{c}\gg r_{g}$ up to the second order of $r_\mathrm{c}/r_g$
because $r_g/r_\mathrm{c}$ approximately coincide with $u_0$
up to the first order of $e$.

\begin{figure}[t]
 \begin{center}
 \includegraphics[keepaspectratio,width=90mm]{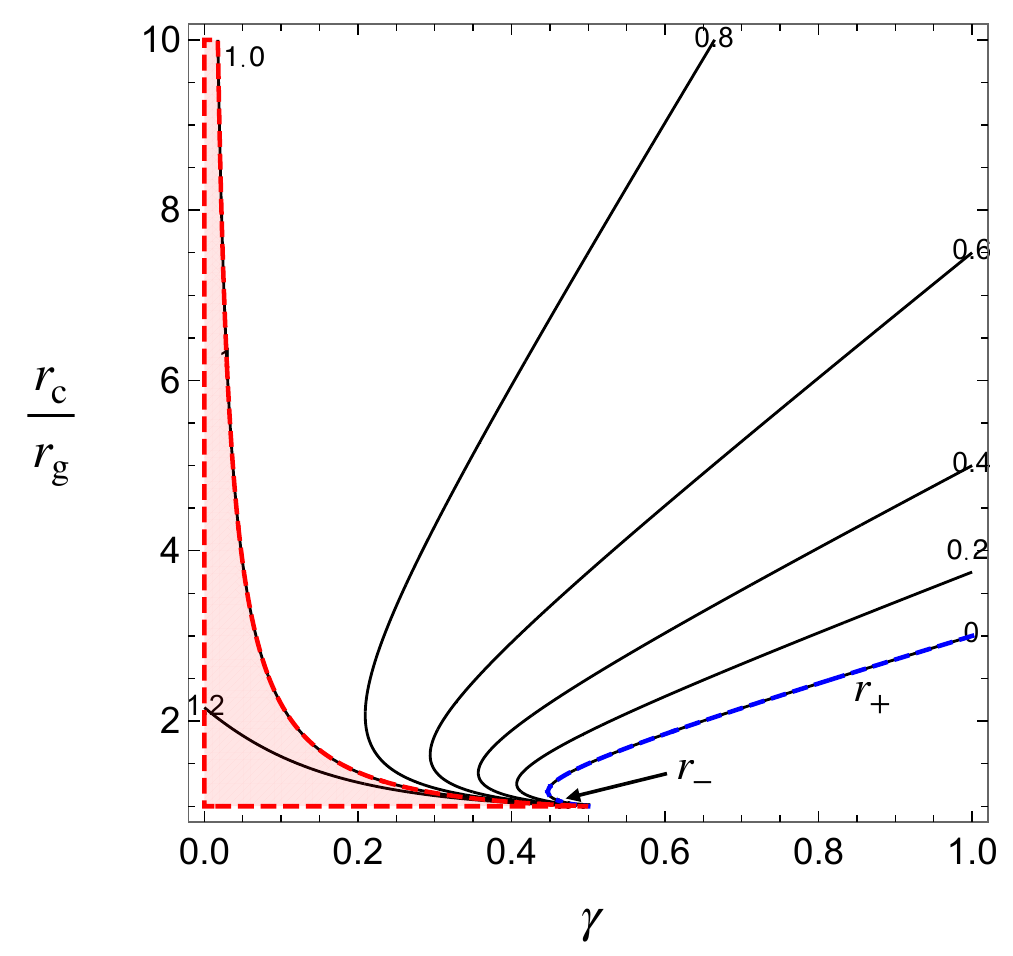}
 \caption{Behavior of the test function $F$.
  The black lines are the contours of $F=\{0,0.2,0.4,0.6,0.8,1.0,1.2\}$.
 The blue dashed curves correspond to $r_\mathrm{\pm}$, respectively. The region enclosed by the red dashed curves corresponds to $F>1$ where the periapsis precession $\Delta_{\text{per}}$ is negative.}
 \label{F1}
 \end{center}
\end{figure}

Let us examine the behavior of the test function~\eqref{testF}.
Figure~\ref{F1} shows the behavior of
$F$ in the $(\gamma,r_\mathrm{c}/r_g)$ contour.
The red region represents the orbits showing the negative periapsis precession.
The behavior of the periapsis precession drastically changes at $\gamma = 1/2$.
For $\gamma \ge 1/2$, the test function $F$ is bounded as  $0<F<1$ and $F(r_+)=0$.
This means that the periapsis precession is always positive
and diverges at  $r_{+}$.

Next, we discuss the $\gamma<1/2$ case.
For $1/\sqrt{5}<\gamma<1/2$, there are two marginally stable circular orbits.
Therefore, the periapsis precession diverges at the two radii.
The periapsis precessions are always positive if $r>r_+$.
On the other hand, the periapsis precession can be negative
in the  $r<r_-$ region.
For $0< \gamma \le 1/\sqrt{5}$, there exist the stable circular orbits with arbitrary radii.
The negative periapsis precession occurs when a test particle moves nearby
the singularity.
As $\gamma$ goes to zero, even if the orbits get away from the singularity, the periapsis precession can be negative.
In short,
the negative periapsis precession occurs only for $0< \gamma<1/2$.
This is the condition that the bound orbits
can move around the vicinity of the singularity.
We can expect that passing close to the singularity is important for the negative periapsis precession.
This expectation is consistent with the previous studies~\cite{Dey:2019fpv,Bambhaniya:2019pbr,Joshi:2019rdo,Dey:2020haf}.

Note that the approximation method performed here can be applied to other spacetimes with naked singularities.
We can reproduce the results obtained in~\cite{Bambhaniya:2019pbr, Joshi:2019rdo}.
It is worth nothing that, in the Reissner-Nordstr\"om spacetime, the negative periapsis precessions can happen.
In fact, for the Reissner-Nordstr\"om spacetime the test function~\eqref{fdef} is given by
\begin{align}
F(r_\mathrm{c}; Q,M)=\frac{-4Q^2+9Q^2Mr_\mathrm{c}-6M^2r_\mathrm{c}^2+Mr_\mathrm{c}^3}{r_\mathrm{c}^2(Mr_\mathrm{c}-Q^2)},
\end{align}
where $Q$ is the charge of the spacetime.
Although this test function depends on $r_{\mathrm{c}}$, roughly, it can be greater than unity for $Q>M$.
Therefore, the negative periapsis precessions occur for the overextremal case.
We can na\" ively expect that a significant change in the behavior of the geodesics is necessary for the periapsis precession to be negative.
Also, we can calculate the periapsis precession
in other spacetimes, e.g., the Kehagias-Sfetsos spacetime\footnote{In~\cite{Kehagias:2009is}, the Kehagias-Sfetsos metric is given by
\begin{align}
d s^{2}= -h(r) d t^{2} + h^{-1}(r) d r^{2} + r^{2}\left(d \theta^{2}+\sin ^{2} \theta d \varphi^{2}\right),\ \ h(r)=1+r^{2} \omega\left[1-\left(1+\frac{4 M}{\omega r^{3}}\right)^{1 / 2}\right],
\end{align}
where $\omega$ is an additional parameter.
This spacetime tends to the Schwarzschild spacetime in $\omega M^2 \to \infty$.
} and the regular black hole spacetimes~\cite{Ayon-Beato:1998hmi, Nicolini:2005vd}.
In these cases, the periapsis precession becomes negative when $\omega<1/2M^2$ and the gravitational source becomes a horizonless object, respectively.
The detailed behaviors of the negative periapsis precession and the comparison between each spacetime are currently under analysis.


\section{Exact expression of the periapsis precession for
\texorpdfstring{$\gamma=1/2$}{g12}\label{pergamma12}}

In this section, we focus on the $\gamma = 1/2$ case,
where we can solve the geodesic equation analytically
in terms of the elliptic function.
There are two advantages to investigate the analytic solution
for the geodesics equation in the FJNWW spacetime with $\gamma = 1/2$.
First, in the previous section, we found that the negative precession does not occur in the
FJNWW spacetime with $\gamma \ge 1/2$ using the small eccentricity approximation.
However, it is not obvious if the negative periapsis precession does not occur in the range where the approximation is not valid.
In order to reveal that, we need to solve the geodesic equation analytically.
If we prove analytically that the negative precession does not occur in the
FJNWW spacetime with $\gamma =1/2$, which is the critical value of occurring the negative precession, for an arbitrary eccentricity, then it would be circumstantial evidence that the negative precession does not occur in the FJNWW spacetime with $\gamma \ge 1/2$.
In fact, in this section, we show that the negative periapsis precession does not occur in the FJNWW spacetime with $\gamma=1/2$ by obtaining the analytic solution to the geodesic equation.
Second, by taking appropriate approximations for the orbits
in the FJNWW spacetime with $\gamma =1/2$,
we can show the validity of the results we have obtained so far
using some approximation methods.

As summarized in the Appendix,
the periapsis precession for $\gamma=1/2$ is given by
\begin{align}
\label{periapsisshiftgamma12}
\Delta_{\text{per}}=4\int_{y_{\text{per}}}^{y_{\text{ap}}}\frac{dy}{\sqrt{P(y)}}-2\pi,
\end{align}
where
\begin{align}
  P(y)\equiv -y^4+2y^2-\lambda y+(\lambda\mu-1),
\end{align}
and we introduced a new variable $y \equiv \sqrt{1-u}$.
Here, $y_{\text{per}}$ and $y_{\text{ap}}$ correspond to the periapsis and the apoapsis of a bound orbit defined as
\begin{align}
y_{\text{per}}\equiv \sqrt{1-u_0(1+e)},\ y_{\text{ap}}\equiv \sqrt{1-u_0(1-e)}.
\end{align}
The polynomial $P(y)$ can be factorized to
\begin{align}
P(y)=-(y-y_{\text{ap}})(y-y_{\text{per}})(y-y_1)(y-y_2).
\end{align}
The parameters $\mu$ and $\lambda$ and the values $y_{\text{per}}, y_{\text{ap}}, y_1$
and $y_2$ are related as
\begin{align}
y_{1,2}&=-\frac{y_{\text{ap}}+y_{\text{per}}}{2}\pm\frac{1}{2}\sqrt{6u_0-2y_{\text{ap}}y_{\text{per}}+2},\\
\lambda&=2u_0(y_{\text{ap}}+y_{\text{per}}),\\
\mu\lambda&=2u_0-u_0^2(1-e^2)+2u_0y_{\text{ap}}y_{\text{per}}.
\end{align}
The bound orbits must satisfy $y_1<y_{\text{per}}$. This condition
is equivalent to the following inequality
\begin{align}
\label{separatrix}
u_0<\frac{2(1+e)}{2(1+e)^2+1},
\end{align}
for bound orbits to exist.
When the eccentricity $e$ is zero, the inequality~\eqref{separatrix}
becomes $u_0<2/3$, which is the condition~\eqref{stableorbitcondition} for $\gamma=1/2$
that is necessary for a stable orbit to exist.

\begin{figure}[t]
 \centering
 \includegraphics[width=110mm]{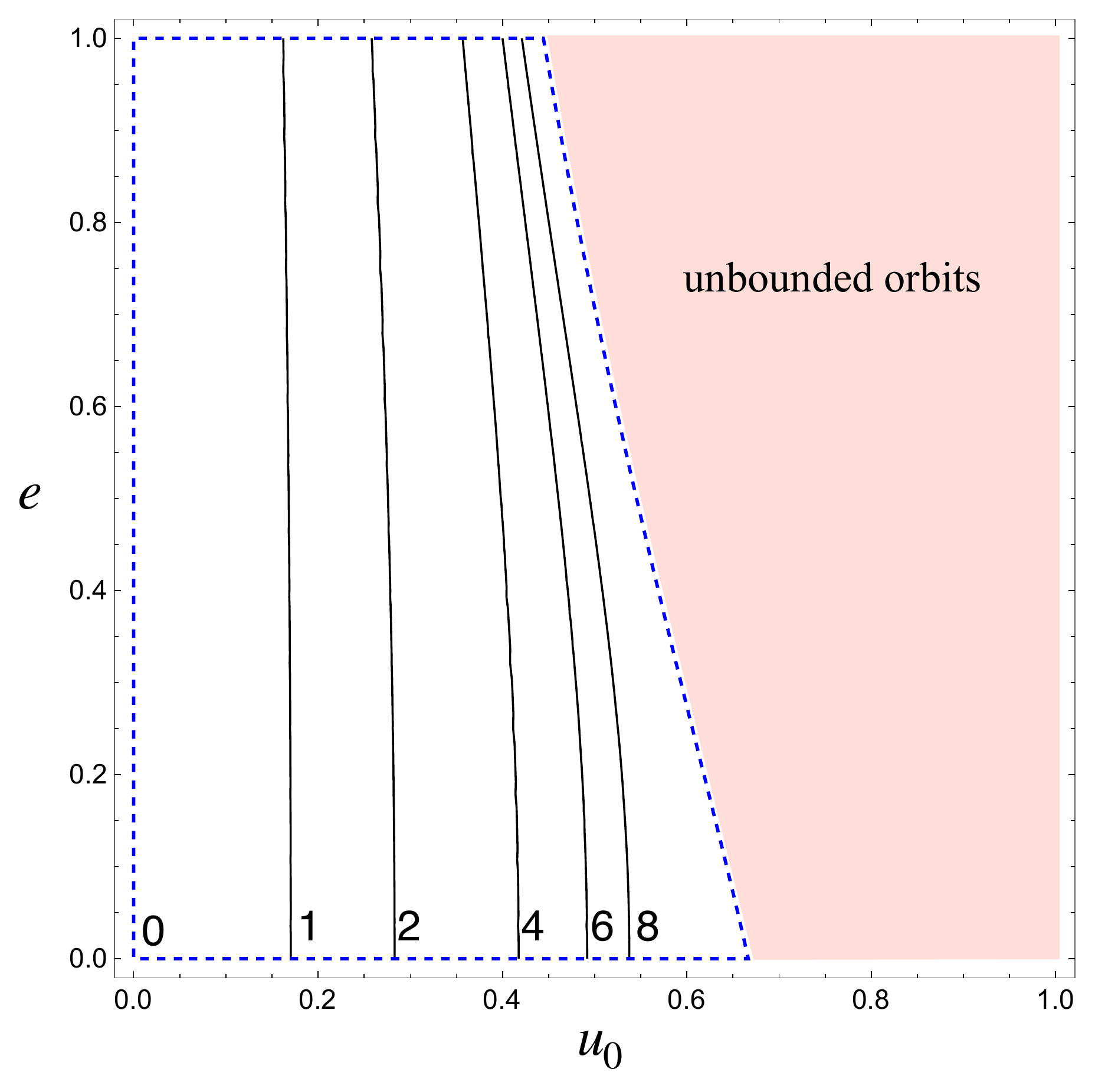}
 \caption{
 Behavior of the periapsis precession $\Delta_{\mathrm{per}}$. The black solid lines are the contours of $\Delta_{\text{per}}=\{0,1,2,4,6,8 \}$. The region inside the blue dashed lines is the bound orbits region. The right boundary of the bounded orbits region corresponds to $u_{0} = 2(1+e)/[2(1+e)^{2}+1]$.  The red shaded region is the unbound orbits region. Note that for $\gamma=1/2$, the periapsis precession is positive for all bound orbits.
 }
 \label{analyticperiapsisgamma12}
\end{figure}

To denote $\Delta_{\text{per}}$ in terms of the complete elliptic integral of the first kind,
we introduce a new variable $y=(u_+\xi + u_-)/(\xi +1)$ where
\begin{align}
u_\pm\equiv\frac{u_0}{y_{\text{ap}}+y_{\text{per}}}\pm\sqrt{\frac{u_0^2}{(y_{\text{ap}}+y_{\text{per}})^2}+y_{\text{ap}}y_{\text{per}}-u_0} .
\end{align}
Then $\Delta_{\text{per}}$ is rewritten as
\begin{align}
\label{ellipticintegral}
\Delta_{\text{per}}=2\frac{u_{+}-u_{-}}{\sqrt{P(u_{+})}}\int_{+\alpha}^{+\infty}\frac{d\xi}{\sqrt{(\xi^2-\alpha^2)(\xi^2-\beta^2)}}-2\pi,
\end{align}
where
\begin{align}
\alpha\equiv
\frac{y_{\text{per}}-u_{-}}{u_{+}-y_{\text{per}}},\
\quad
\beta\equiv
\frac{y_{1}-u_{-}}{u_{+}-y_{1}}.
\end{align}
We note that $\alpha^2>\beta^2>0$.
Performing a variable transformation as $\xi=\alpha/z$,
we finally obtain the exact expression of the periapsis precession
in terms of  the complete elliptic integral of the first kind as
\begin{align}
  \Delta_{\mathrm{per}}
  &=
  2 \frac{u_{+} - u_{-}}{\sqrt{(u_{-}-y_{\text{ap}})(u_{-}-y_{\text{per}})(u_+-y_1)(u_+-y_2)}} \int ^{1}_{0} \frac{dz}{\sqrt{(1-z^{2})(1- \tilde{k} z^{2})}} -2\pi  \nonumber \\
&=8\frac{u_{+}-u_{-}}{\sqrt{(u_{-}-y_{\text{ap}})(u_{-}-y_{\text{per}})(u_+-y_1)(u_+-y_2)}}K(\tilde{k})-2\pi,
  \label{periapsisshiftanalytic12}
\end{align}
where the parameter $\tilde{k}$ is
\begin{align}
\tilde{k} \equiv \left[\frac{(y_{1}-u_{-})(u_{+}-y_{\text{per}})}{(u_{+}-y_{1})(y_{\text{per}}-u_{-})} \right]^{2},
\end{align}
and the range of $\tilde{k}$ is given by $0< \tilde{k} <1$.
Figure~\ref{analyticperiapsisgamma12} shows
the behavior of $ \Delta_{\mathrm{per}}$ in the $(u_0,e)$ diagram.
The bound orbits physically acceptable can exist only in the left region
divided by the blue dashed line, which corresponds to $u_0 = 2(1+e)/[2(1+e)^2+1]$.
We see that the periapsis precession is positive  in the entire parameter region
for $\gamma=1/2$.
We note that the periapsis precession becomes larger as $u_{0}$ and $e$ become larger.
This implies that, for the $\gamma=1/2$ case, the orbits with the small semilatus have the large precession.
This means that even if an orbit gets close to the singularity, the value of the precession always does not become smaller.
Since the scalar charge is the source of gravity as well as the cause of the drastic change of the geodesic structure, the value of the precession is determined by these two effects.

In order to compare to the approximate solution~\eqref{periapsisshift21app},
we expand Eq.~\eqref{periapsisshiftanalytic12} up to the second order of $u_0$ as
\begin{align}
\label{periapsisshiftanalytic12app}
\Delta_{\text{per}}=
   \frac{3}{2}\pi u_0
 + \frac{3}{16}(9+2e^2)\pi u_0^2
 + \frac{27}{64} (5+3 e^{2})\pi u_{0}^{3}
 + \mathcal{O}(u_0^{4}).
\end{align}
This result is consistent with Eq.~\eqref{periapsisshift21app} for $\gamma=1/2$.
Furthermore, we verify that the approximate periapsis precession~\eqref{geodevper} is consistent with Eq.~\eqref{periapsisshiftanalytic12} if $e\ll 1$.
Expanding Eq.~\eqref{periapsisshiftanalytic12}, we obtain
\begin{align}
\label{periapsisshiftanalytic12appe}
\Delta_{\text{per}}
=
2\pi
\left(
\frac{1}{\sqrt{1-3u_{0}/2}}-1
\right)
+\mathcal{O}(e^2).
\end{align}
This result is also consistent with Eq.~\eqref{geodevper} for $\gamma=1/2$.

As shown in the Appendix, for some particular values of $\gamma$,
the analytical solution can be obtained by the elliptic function,
which is expected to be analyzed for the more general case of $\gamma$ by using hyperelliptic functions~\cite{Hackmann:2008zz,Enolski:2010if}.

\section{Summary and Discussion}
In this paper, we studied the timelike geodesics and evaluated the periapsis precession in the FJNWW spacetime under the several approximations.
We complemented the result of~\cite{Joshi:2019rdo} for the FJNWW spacetime under the weak field and the small eccentricity approximation.
We showed that the negative periapsis precessions occur when the spacetime sufficiently deviates from the Schwarzschild spacetime: the parameter $\gamma$ is small enough.
Next we suggested two approximation methods to relax the approximation
conditions used in~\cite{Joshi:2019rdo}.
First, employing only the weak field approximation $(u_{0} \ll 1)$, we obtained the analytical solution of the approximate geodesic equation~\eqref{jnwweakeq1}.
This allowed us to find the periapsis precession for orbits
with arbitrary eccentricity.
Second, we assumed only small eccentricity $(e \ll 1)$
and solved the geodesic deviation equation around the circular orbit.
In this case, we showed that the negative periapsis precession can occur
when $\gamma$ is smaller than $1/2$, where particles can move around the vicinity of the singularity.
This implies that it is important that a particle moves around close enough to the singularity for the periapsis precession to be negative.
Our results indicate that a naked singularity spacetime may exhibit the characteristic behaviors in the periapsis precession.
Although with the observations of the periapsis precession, it is difficult to distinguish the FJNWW spacetime from other naked singularity spacetimes, it is possible to test the models of naked singularity spacetimes as have been done in~\cite{Dey:2019fpv}.

Furthermore, we found that the analytical expression of the periapsis precession in terms of the elliptic integral for $\gamma = 1/2$.
We saw that for $\gamma=1/2$, the negative periapsis precession never occurs: $\Delta_{\mathrm{per}}$ is always positive.
We confirmed that the approximate solutions are consistent with the analytical solution for at least $\gamma=1/2$.
As shown in the Appendix, for some particular values of $\gamma$,
the analytical solution can be obtained by the elliptic function,
which is expected to be analyzed for the more general case of $\gamma$ by using hyperelliptic functions~\cite{Hackmann:2008zz,Enolski:2010if}.
The investigation and utilization of such analytic solutions
are left for a future work.

\appendix
\section{Analytical solution for geodesic equation\label{ana}}
We can solve the geodesic equation~\eqref{geodesiceq1}
in the FJNWW spacetime following~\cite{Enolski:2010if}.
Introducing $y=(1-r_g/r)^{1/\alpha}$ and
substituting it into Eq.~\eqref{geodesiceq1},
we obtain
\begin{align}
\label{hyperdeffeq}
\alpha^2\left(\frac{dy}{d\phi}\right)^2=\mu\lambda y^{2(1-\alpha\gamma)}-\epsilon\lambda y^{2-\alpha\gamma}-y^{2-\alpha}+2y^2-y^{2+\alpha}.
\end{align}
The solution of Eq.~\eqref{geodesiceq1} is represented
by the elliptic function if Eq.~\eqref{geodesiceq1} is
reduced to the following form
\begin{align}
\left(\frac{dy}{dx}\right)^2=P_{3,4}(y),
\end{align}
where $P_{3,4}(y)$ is a third or fourth order polynomial.
This is realized for $(\alpha,\gamma)=(\forall \alpha,0), (2,1/2), (2,1/4)$.
In this appendix, we give analytical solutions
for each case.
Note that for the third case $(2,1/4)$,
an analytic solution for a null geodesics ($\epsilon=0$) can be found.

\subsection{\texorpdfstring{Timelike and null geodesics for $\gamma=0$}{g0}}

\begin{figure}[t]
 \centering
\includegraphics[width=150mm]{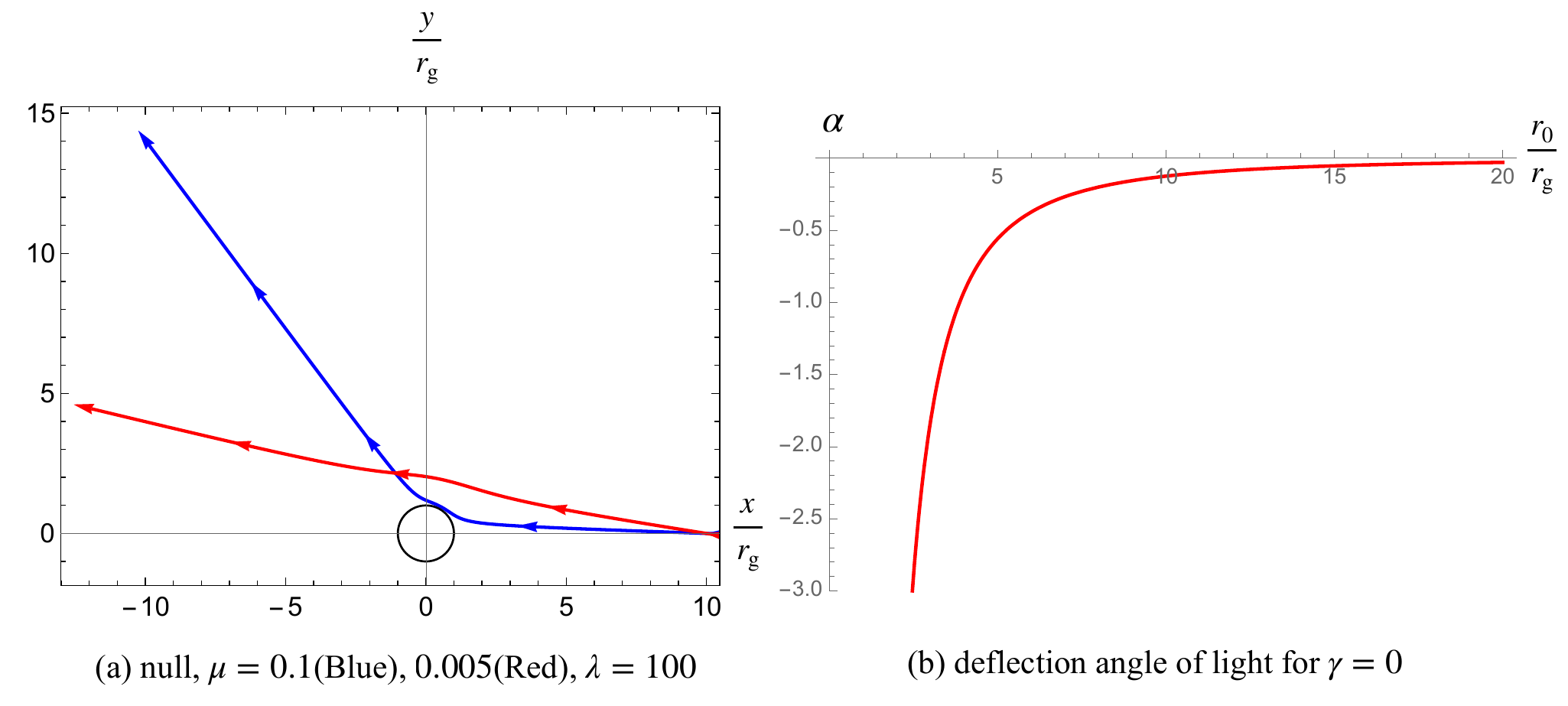}
 \caption{Left panel (a) : the null escape orbit for $\gamma=0$. $\mu=0.1$(blue), 0.005(red), $\lambda=100$.
 The timelike geodesics also have the same orbit when we choose an appropriate value of $A$.
 Right panel (b) : the deflection angle of light. For $\gamma = 0$, all of the orbits have negative deflection angles.}
 \label{gamma0}
\end{figure}

When $\gamma=0$, we can obtain an analytic solution of
the geodesic equation for arbitrary $\alpha$, so we set here $\alpha=1$.
Then Eq.~\eqref{geodesiceq1} reduces to the following differential equation:
\begin{align}
\label{Elliptic}
\left(\frac{dy}{d\phi}\right)^2=-y^3+A y^2-y,
\end{align}
where $A\equiv\lambda(\mu-\epsilon)+2$.
We note that for $\gamma=0$, the difference between the timelike geodesics ($\epsilon=1$) and the null geodesics ($\epsilon=0$) appears as the constant shift of $A$.
That is, both the timelike geodesics and the null geodesics are represented by the same equation~\eqref{Elliptic}.
With a substitution $y=-4z+A/3$, Eq.~\eqref{Elliptic} is rewritten as
\begin{align}
\left(\frac{dz}{d\phi}\right)^2=4z^3-g_2z-g_3,
\end{align}
where
\begin{align}
&g_2=\frac{1}{4}\left(\frac{1}{3}A^2-1\right), \\
& g_3=\frac{1}{48}A\left(1-\frac{2}{9}A^2\right).
\end{align}
This equation has a solution represented by the Weierstrass function as
\begin{align}
z(\phi)=\wp(\phi-\phi_\mathrm{in};g_2,g_3),
\end{align}
where
\begin{align}
\phi_\mathrm{in}=\phi_0+\int_{z_0}^\infty\frac{dz}{\sqrt{4z^3-g_2z-g_3}},
\quad
z_0=\frac{1}{4}\left(\frac{r_g}{r_0}-1+\frac{1}{3}A\right).
\end{align}
Finally we find the analytical solution of the geodesic equation in the FJNWW spacetime with $\gamma=0$ as
\begin{align}
r(\phi)=r_g\left[4\wp(\phi-\phi_\mathrm{in})+1-\frac{1}{3}A\right]^{-1}.
\end{align}
Figure~\ref{gamma0} shows the null escape orbits. In this spacetime, all orbits have the negative deflection angles.

\subsection{\texorpdfstring{Timelike and null geodesics for $\gamma=1/2$}{g12}}

\begin{figure}[t]
 \centering
\includegraphics[width=150mm]{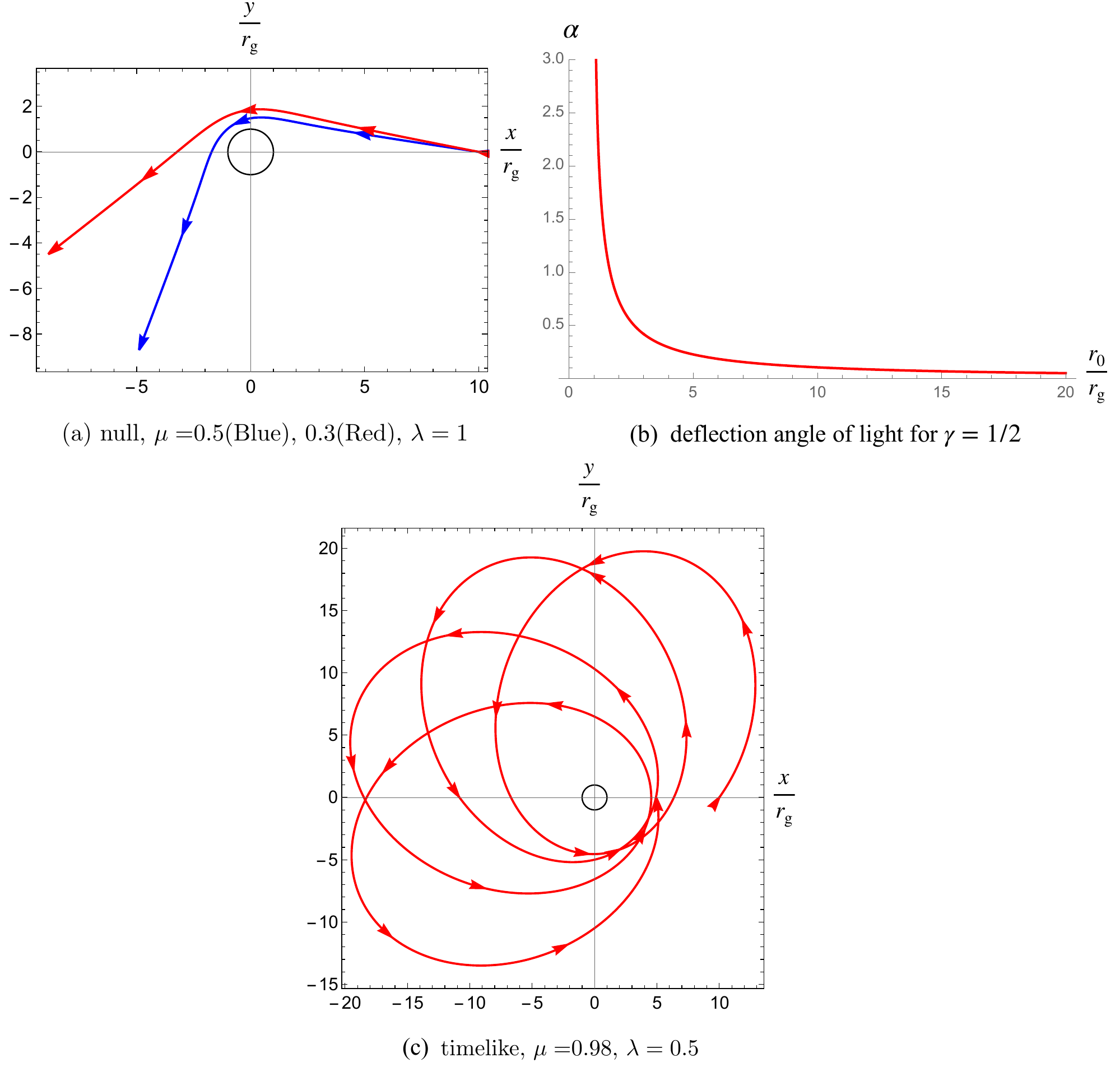}
 \caption{Top left panel (a) : the null escape orbits for $\gamma=1/2$. Top right panel (b) : the deflection angle of light. For $\gamma=1/2$, the deflection angle of light is always positive. Bottom panel (c) : the timelike bound orbit for $\gamma=1/2$. For $\gamma=1/2$, the periapsis precession is always positive.}
 \label{gamma12}
\end{figure}

Equation~\eqref{hyperdeffeq} with $\alpha =2$ and $\gamma = 1/2$ reduces to
\begin{align}
\label{diff1}
\left(\frac{dy}{d\phi}\right)^2=-\frac{1}{4}y^4+\frac{1}{2}y^2-\epsilon\frac{\lambda}{4}y+\frac{1}{4}(\lambda\mu-1)=Y_{1/2}(y).
\end{align}
Substituting $y=\xi^{-1}+y_\mathrm{zero}$,
where $y_\mathrm{zero}$ is the algebraic roots of $Y_{1/2}(y)$, we obtain
\begin{align}
\left(\frac{d\xi}{d\phi}\right)^2=\sum_{j=0}^3a_j\xi^j,\ \ a_j=\frac{1}{(4-j)!}\left.\frac{d^{4-j}Y_{1/2}(y)}{dy^{4-j}}\right|_{y=y_\mathrm{zero}}.
\end{align}
Applying a transformation as
$\xi=a_3^{-1}(4z-a_2/3)$,
we obtain the standard form of the elliptic function
\begin{align}
\label{gamma12diff}
\left(\frac{dz}{d\phi}\right)^2=4z^3-g_2z-g_3,
\end{align}
where
\begin{align}
&g_2=\frac{1}{4}\left(\frac{1}{3}a_2^2-a_1a_3\right), \\
&
g_3=\frac{1}{16}\left(\frac{1}{3}a_1a_2a_3-\frac{2}{27}a_2^3-a_0a_3^2\right).
\end{align}
Finally we write the solution of the geodesic equation~\eqref{geodesiceq1}
with $\gamma = 1/2$ as
\begin{align}
r(\phi)=r_g\left[1-\left(\frac{a_3}{4\wp(\phi-\phi_\mathrm{in})-\frac{a_2}{3}} + y_\mathrm{zero}\right)^2 \right]^{-1},
\end{align}
where
\begin{align}
\phi_\mathrm{in}=\phi_0+\int_{z_0}^\infty\frac{dz}{\sqrt{4z^3-g_2z-g_3}},
\quad
z_0=\frac{1}{4}\left[\frac{a_3}{(1-\frac{r_g}{r_0})^{\frac{1}{2}}-y_\mathrm{zero}}+\frac{a_2}{3}\right].
\end{align}
Figure~\ref{gamma12} shows null escape orbits and a timelike bound orbit.
Using this analytical solution, we can see that there are only the positive periapsis precession and the positive deflection angle.

\begin{figure}[t]
 \centering
\includegraphics[width=150mm]{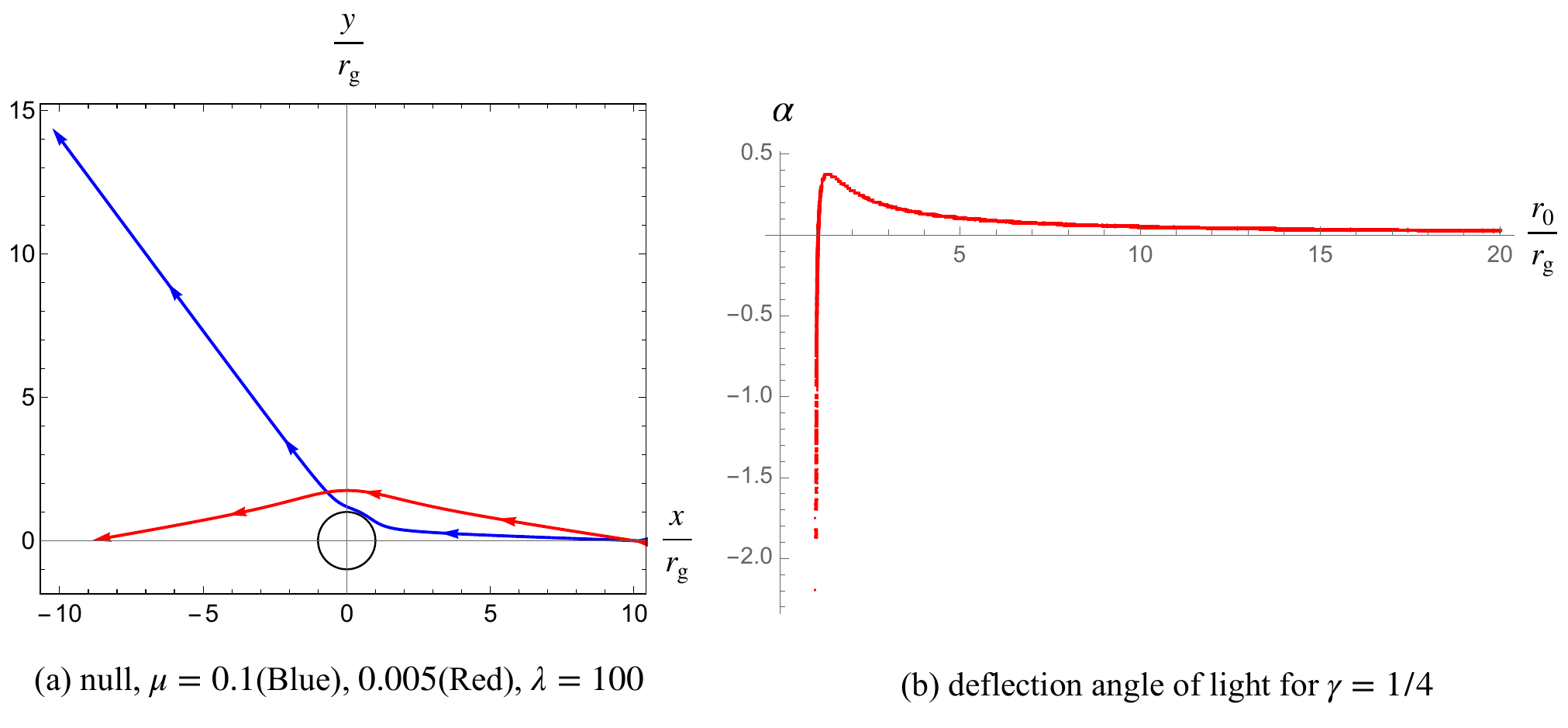}
\caption{Left panel (a) : the null escape orbit for $\gamma=1/4$.
$\mu=$0.1(blue), 0.005(red), $\lambda=100$.
Right panel (b) : the deflection angle of light. For $\gamma=1/4$, when an orbit passes the close of the singularity, the deflection angle of light becomes negative.   }
 \label{gamma14}
\end{figure}

\subsection{\texorpdfstring{Null geodesics for  $\gamma=1/4$}{g14} (null)}
The geodesic equation with $\gamma=1/4$  can be also solved
for a null geodesics exactly.
Equation~\eqref{hyperdeffeq} with $\alpha =2$ and $\gamma = 1/4$
can be transformed as
\begin{align}
\label{diff2}
\left(\frac{dy}{d\phi}\right)^2=-\frac{1}{4}y^4+\frac{1}{2}y^2+\frac{\mu\lambda}{4}y-\frac{1}{4}=Y_{1/4}(y).
\end{align}
Using  the algebraic roots of $Y_{1/4}=0$, $y_\mathrm{zero}$,
we introduce $y=\xi^{-1}+y_\mathrm{zero}$
and substitute it into Eq.~\eqref{diff2}, we obtain
\begin{align}
\label{diff3}
\left(\frac{d\xi}{d\phi}\right)^2=\sum_{j=0}^3b_j\xi^j,\ \ b_j=\frac{1}{(4-j)!}\left.\frac{d^{4-j}Y_{1/4}(y)}{dy^{4-j}}\right|_{y=y_\mathrm{zero}}.
\end{align}
An additional substitution $\xi=b_3^{-1}(4z-b_2/3)$ transforms Eq.~\eqref{diff3} into
\begin{align}
\left(\frac{dz}{d\phi}\right)^2=4z^3-g_2z-g_3,
\end{align}
where
\begin{align}
&g_2=\frac{1}{4}\left(\frac{1}{3}b_2^2-b_1b_3\right),\\
&g_3=\frac{1}{16}\left(\frac{1}{3}b_1b_2b_3-\frac{2}{27}b_2^3b_0b_3^2\right).
\end{align}
Finally the analytical solution for the geodesic equation for $\gamma = 1/4$
is found as
\begin{align}
r(\phi)=r_g\left[1-\left(\frac{b_3}{4\wp(\phi-\phi_\mathrm{in})-\frac{b_2}{3}}+y_\mathrm{zero}\right)^2\right]^{-1},
\end{align}
where
\begin{align}
\phi_\mathrm{in}=\phi_0+\int_{z_0}^\infty\frac{dz}{\sqrt{4z^3-g_2z-g_3}},\quad
z_0=\frac{1}{4}\left[\frac{b_3}{(1-\frac{r_g}{r_0})^{\frac{1}{2}}-y_\mathrm{zero}}+\frac{b_2}{3}\right].
\end{align}
Figure~\ref{gamma14} shows null escape orbits. The blue and red curves have negative and positive deflection angles, respectively.

\section*{Acknowledgments}
We would like to thank to T. Harada and M. Kimura for helpful discussions.

\bibliographystyle{apsrev4-2}
\bibliography{refpaper}

\begin{thebibliography}{39}%
\makeatletter
\providecommand \@ifxundefined [1]{%
 \@ifx{#1\undefined}
}%
\providecommand \@ifnum [1]{%
 \ifnum #1\expandafter \@firstoftwo
 \else \expandafter \@secondoftwo
 \fi
}%
\providecommand \@ifx [1]{%
 \ifx #1\expandafter \@firstoftwo
 \else \expandafter \@secondoftwo
 \fi
}%
\providecommand \natexlab [1]{#1}%
\providecommand \enquote  [1]{``#1''}%
\providecommand \bibnamefont  [1]{#1}%
\providecommand \bibfnamefont [1]{#1}%
\providecommand \citenamefont [1]{#1}%
\providecommand \href@noop [0]{\@secondoftwo}%
\providecommand \href [0]{\begingroup \@sanitize@url \@href}%
\providecommand \@href[1]{\@@startlink{#1}\@@href}%
\providecommand \@@href[1]{\endgroup#1\@@endlink}%
\providecommand \@sanitize@url [0]{\catcode `\\12\catcode `\$12\catcode
  `\&12\catcode `\#12\catcode `\^12\catcode `\_12\catcode `\%12\relax}%
\providecommand \@@startlink[1]{}%
\providecommand \@@endlink[0]{}%
\providecommand \url  [0]{\begingroup\@sanitize@url \@url }%
\providecommand \@url [1]{\endgroup\@href {#1}{\urlprefix }}%
\providecommand \urlprefix  [0]{URL }%
\providecommand \Eprint [0]{\href }%
\providecommand \doibase [0]{https://doi.org/}%
\providecommand \selectlanguage [0]{\@gobble}%
\providecommand \bibinfo  [0]{\@secondoftwo}%
\providecommand \bibfield  [0]{\@secondoftwo}%
\providecommand \translation [1]{[#1]}%
\providecommand \BibitemOpen [0]{}%
\providecommand \bibitemStop [0]{}%
\providecommand \bibitemNoStop [0]{.\EOS\space}%
\providecommand \EOS [0]{\spacefactor3000\relax}%
\providecommand \BibitemShut  [1]{\csname bibitem#1\endcsname}%
\let\auto@bib@innerbib\@empty
\bibitem [{\citenamefont {Penrose}(1965)}]{PhysRevLett.14.57}%
  \BibitemOpen
  \bibfield  {author} {\bibinfo {author} {\bibfnamefont {R.}~\bibnamefont
  {Penrose}},\ }\href {https://doi.org/10.1103/PhysRevLett.14.57} {\bibfield
  {journal} {\bibinfo  {journal} {Phys. Rev. Lett.}\ }\textbf {\bibinfo
  {volume} {14}},\ \bibinfo {pages} {57} (\bibinfo {year} {1965})}\BibitemShut
  {NoStop}%
\bibitem [{\citenamefont {Hawking}(1967)}]{Hawking:1967ju}%
  \BibitemOpen
  \bibfield  {author} {\bibinfo {author} {\bibfnamefont {S.}~\bibnamefont
  {Hawking}},\ }\href {https://doi.org/10.1098/rspa.1967.0164} {\bibfield
  {journal} {\bibinfo  {journal} {Proc. Roy. Soc. Lond. A}\ }\textbf {\bibinfo
  {volume} {300}},\ \bibinfo {pages} {187} (\bibinfo {year}
  {1967})}\BibitemShut {NoStop}%
\bibitem [{\citenamefont {Hawking}\ and\ \citenamefont
  {Penrose}(1970)}]{Hawking:1970zqf}%
  \BibitemOpen
  \bibfield  {author} {\bibinfo {author} {\bibfnamefont {S.~W.}\ \bibnamefont
  {Hawking}}\ and\ \bibinfo {author} {\bibfnamefont {R.}~\bibnamefont
  {Penrose}},\ }\href {https://doi.org/10.1098/rspa.1970.0021} {\bibfield
  {journal} {\bibinfo  {journal} {Proc. Roy. Soc. Lond. A}\ }\textbf {\bibinfo
  {volume} {314}},\ \bibinfo {pages} {529} (\bibinfo {year}
  {1970})}\BibitemShut {NoStop}%
\bibitem [{\citenamefont {Penrose}(1969)}]{Penrose:1969pc}%
  \BibitemOpen
  \bibfield  {author} {\bibinfo {author} {\bibfnamefont {R.}~\bibnamefont
  {Penrose}},\ }\href {https://doi.org/10.1023/A:1016578408204} {\bibfield
  {journal} {\bibinfo  {journal} {Riv. Nuovo Cim.}\ }\textbf {\bibinfo {volume}
  {1}},\ \bibinfo {pages} {252} (\bibinfo {year} {1969})}\BibitemShut {NoStop}%
\bibitem [{\citenamefont {{Penrose}}(1979)}]{1979grec.conf..581P}%
  \BibitemOpen
  \bibfield  {author} {\bibinfo {author} {\bibfnamefont {R.}~\bibnamefont
  {{Penrose}}},\ }in\ \href@noop {} {\emph {\bibinfo {booktitle} {General
  Relativity: An Einstein centenary survey}}},\ \bibinfo {editor} {edited by\
  \bibinfo {editor} {\bibfnamefont {S.~W.}\ \bibnamefont {{Hawking}}}\ and\
  \bibinfo {editor} {\bibfnamefont {W.}~\bibnamefont {{Israel}}}}\ (\bibinfo
  {year} {1979})\ pp.\ \bibinfo {pages} {581--638}\BibitemShut {NoStop}%
\bibitem [{\citenamefont {Harada}\ \emph {et~al.}(2002)\citenamefont {Harada},
  \citenamefont {Iguchi},\ and\ \citenamefont {Nakao}}]{Harada:2001nj}%
  \BibitemOpen
  \bibfield  {author} {\bibinfo {author} {\bibfnamefont {T.}~\bibnamefont
  {Harada}}, \bibinfo {author} {\bibfnamefont {H.}~\bibnamefont {Iguchi}},\
  and\ \bibinfo {author} {\bibfnamefont {K.-i.}\ \bibnamefont {Nakao}},\ }\href
  {https://doi.org/10.1143/PTP.107.449} {\bibfield  {journal} {\bibinfo
  {journal} {Prog. Theor. Phys.}\ }\textbf {\bibinfo {volume} {107}},\ \bibinfo
  {pages} {449} (\bibinfo {year} {2002})},\ \Eprint
  {https://arxiv.org/abs/gr-qc/0204008} {arXiv:gr-qc/0204008} \BibitemShut
  {NoStop}%
\bibitem [{\citenamefont {Joshi}\ and\ \citenamefont
  {Malafarina}(2011)}]{Joshi:2011rlc}%
  \BibitemOpen
  \bibfield  {author} {\bibinfo {author} {\bibfnamefont {P.~S.}\ \bibnamefont
  {Joshi}}\ and\ \bibinfo {author} {\bibfnamefont {D.}~\bibnamefont
  {Malafarina}},\ }\href {https://doi.org/10.1142/S0218271811020792} {\bibfield
   {journal} {\bibinfo  {journal} {Int. J. Mod. Phys. D}\ }\textbf {\bibinfo
  {volume} {20}},\ \bibinfo {pages} {2641} (\bibinfo {year} {2011})},\ \Eprint
  {https://arxiv.org/abs/1201.3660} {arXiv:1201.3660 [gr-qc]} \BibitemShut
  {NoStop}%
\bibitem [{\citenamefont {Fisher}(1948)}]{Fisher:1948yn}%
  \BibitemOpen
  \bibfield  {author} {\bibinfo {author} {\bibfnamefont {I.}~\bibnamefont
  {Fisher}},\ }\href@noop {} {\bibfield  {journal} {\bibinfo  {journal} {Zh.
  Eksp. Teor. Fiz.}\ }\textbf {\bibinfo {volume} {18}},\ \bibinfo {pages} {636}
  (\bibinfo {year} {1948})},\ \Eprint {https://arxiv.org/abs/gr-qc/9911008}
  {arXiv:gr-qc/9911008} \BibitemShut {NoStop}%
\bibitem [{\citenamefont {Janis}\ \emph {et~al.}(1968)\citenamefont {Janis},
  \citenamefont {Newman},\ and\ \citenamefont {Winicour}}]{Janis:1968zz}%
  \BibitemOpen
  \bibfield  {author} {\bibinfo {author} {\bibfnamefont {A.~I.}\ \bibnamefont
  {Janis}}, \bibinfo {author} {\bibfnamefont {E.~T.}\ \bibnamefont {Newman}},\
  and\ \bibinfo {author} {\bibfnamefont {J.}~\bibnamefont {Winicour}},\ }\href
  {https://doi.org/10.1103/PhysRevLett.20.878} {\bibfield  {journal} {\bibinfo
  {journal} {Phys. Rev. Lett.}\ }\textbf {\bibinfo {volume} {20}},\ \bibinfo
  {pages} {878} (\bibinfo {year} {1968})}\BibitemShut {NoStop}%
\bibitem [{\citenamefont {Wyman}(1981)}]{Wyman:1981bd}%
  \BibitemOpen
  \bibfield  {author} {\bibinfo {author} {\bibfnamefont {M.}~\bibnamefont
  {Wyman}},\ }\href {https://doi.org/10.1103/PhysRevD.24.839} {\bibfield
  {journal} {\bibinfo  {journal} {Phys. Rev. D}\ }\textbf {\bibinfo {volume}
  {24}},\ \bibinfo {pages} {839} (\bibinfo {year} {1981})}\BibitemShut
  {NoStop}%
\bibitem [{\citenamefont {Virbhadra}(1997)}]{Virbhadra:1997ie}%
  \BibitemOpen
  \bibfield  {author} {\bibinfo {author} {\bibfnamefont {K.~S.}\ \bibnamefont
  {Virbhadra}},\ }\href {https://doi.org/10.1142/S0217751X97002577} {\bibfield
  {journal} {\bibinfo  {journal} {Int. J. Mod. Phys. A}\ }\textbf {\bibinfo
  {volume} {12}},\ \bibinfo {pages} {4831} (\bibinfo {year} {1997})},\ \Eprint
  {https://arxiv.org/abs/gr-qc/9701021} {arXiv:gr-qc/9701021} \BibitemShut
  {NoStop}%
\bibitem [{\citenamefont {Roberts}(1993)}]{Roberts:1993re}%
  \BibitemOpen
  \bibfield  {author} {\bibinfo {author} {\bibfnamefont {M.~D.}\ \bibnamefont
  {Roberts}},\ }\href {https://doi.org/10.1007/BF00627140} {\bibfield
  {journal} {\bibinfo  {journal} {Astrophys. Space Sci.}\ }\textbf {\bibinfo
  {volume} {200}},\ \bibinfo {pages} {331} (\bibinfo {year}
  {1993})}\BibitemShut {NoStop}%
\bibitem [{\citenamefont {Xanthopoulos}\ and\ \citenamefont
  {Zannias}(1989)}]{Xanthopoulos:1989kb}%
  \BibitemOpen
  \bibfield  {author} {\bibinfo {author} {\bibfnamefont {B.~C.}\ \bibnamefont
  {Xanthopoulos}}\ and\ \bibinfo {author} {\bibfnamefont {T.}~\bibnamefont
  {Zannias}},\ }\href {https://doi.org/10.1103/PhysRevD.40.2564} {\bibfield
  {journal} {\bibinfo  {journal} {Phys. Rev. D}\ }\textbf {\bibinfo {volume}
  {40}},\ \bibinfo {pages} {2564} (\bibinfo {year} {1989})}\BibitemShut
  {NoStop}%
\bibitem [{\citenamefont {Virbhadra}\ \emph {et~al.}(1998)\citenamefont
  {Virbhadra}, \citenamefont {Narasimha},\ and\ \citenamefont
  {Chitre}}]{Virbhadra:1998dy}%
  \BibitemOpen
  \bibfield  {author} {\bibinfo {author} {\bibfnamefont {K.~S.}\ \bibnamefont
  {Virbhadra}}, \bibinfo {author} {\bibfnamefont {D.}~\bibnamefont
  {Narasimha}},\ and\ \bibinfo {author} {\bibfnamefont {S.~M.}\ \bibnamefont
  {Chitre}},\ }\href@noop {} {\bibfield  {journal} {\bibinfo  {journal}
  {Astron. Astrophys.}\ }\textbf {\bibinfo {volume} {337}},\ \bibinfo {pages}
  {1} (\bibinfo {year} {1998})},\ \Eprint
  {https://arxiv.org/abs/astro-ph/9801174} {arXiv:astro-ph/9801174}
  \BibitemShut {NoStop}%
\bibitem [{\citenamefont {Gyulchev}\ \emph {et~al.}(2019)\citenamefont
  {Gyulchev}, \citenamefont {Nedkova}, \citenamefont {Vetsov},\ and\
  \citenamefont {Yazadjiev}}]{Gyulchev:2019tvk}%
  \BibitemOpen
  \bibfield  {author} {\bibinfo {author} {\bibfnamefont {G.}~\bibnamefont
  {Gyulchev}}, \bibinfo {author} {\bibfnamefont {P.}~\bibnamefont {Nedkova}},
  \bibinfo {author} {\bibfnamefont {T.}~\bibnamefont {Vetsov}},\ and\ \bibinfo
  {author} {\bibfnamefont {S.}~\bibnamefont {Yazadjiev}},\ }\href
  {https://doi.org/10.1103/PhysRevD.100.024055} {\bibfield  {journal} {\bibinfo
   {journal} {Phys. Rev. D}\ }\textbf {\bibinfo {volume} {100}},\ \bibinfo
  {pages} {024055} (\bibinfo {year} {2019})},\ \Eprint
  {https://arxiv.org/abs/1905.05273} {arXiv:1905.05273 [gr-qc]} \BibitemShut
  {NoStop}%
\bibitem [{\citenamefont {Sau}\ \emph {et~al.}(2020)\citenamefont {Sau},
  \citenamefont {Banerjee},\ and\ \citenamefont {SenGupta}}]{Sau:2020xau}%
  \BibitemOpen
  \bibfield  {author} {\bibinfo {author} {\bibfnamefont {S.}~\bibnamefont
  {Sau}}, \bibinfo {author} {\bibfnamefont {I.}~\bibnamefont {Banerjee}},\ and\
  \bibinfo {author} {\bibfnamefont {S.}~\bibnamefont {SenGupta}},\ }\href
  {https://doi.org/10.1103/PhysRevD.102.064027} {\bibfield  {journal} {\bibinfo
   {journal} {Phys. Rev. D}\ }\textbf {\bibinfo {volume} {102}},\ \bibinfo
  {pages} {064027} (\bibinfo {year} {2020})},\ \Eprint
  {https://arxiv.org/abs/2004.02840} {arXiv:2004.02840 [gr-qc]} \BibitemShut
  {NoStop}%
\bibitem [{\citenamefont {Chowdhury}\ \emph {et~al.}(2012)\citenamefont
  {Chowdhury}, \citenamefont {Patil}, \citenamefont {Malafarina},\ and\
  \citenamefont {Joshi}}]{Chowdhury:2011aa}%
  \BibitemOpen
  \bibfield  {author} {\bibinfo {author} {\bibfnamefont {A.~N.}\ \bibnamefont
  {Chowdhury}}, \bibinfo {author} {\bibfnamefont {M.}~\bibnamefont {Patil}},
  \bibinfo {author} {\bibfnamefont {D.}~\bibnamefont {Malafarina}},\ and\
  \bibinfo {author} {\bibfnamefont {P.~S.}\ \bibnamefont {Joshi}},\ }\href
  {https://doi.org/10.1103/PhysRevD.85.104031} {\bibfield  {journal} {\bibinfo
  {journal} {Phys. Rev. D}\ }\textbf {\bibinfo {volume} {85}},\ \bibinfo
  {pages} {104031} (\bibinfo {year} {2012})},\ \Eprint
  {https://arxiv.org/abs/1112.2522} {arXiv:1112.2522 [gr-qc]} \BibitemShut
  {NoStop}%
\bibitem [{\citenamefont {Dey}\ \emph {et~al.}(2019)\citenamefont {Dey},
  \citenamefont {Joshi}, \citenamefont {Joshi},\ and\ \citenamefont
  {Bambhaniya}}]{Dey:2019fpv}%
  \BibitemOpen
  \bibfield  {author} {\bibinfo {author} {\bibfnamefont {D.}~\bibnamefont
  {Dey}}, \bibinfo {author} {\bibfnamefont {P.~S.}\ \bibnamefont {Joshi}},
  \bibinfo {author} {\bibfnamefont {A.}~\bibnamefont {Joshi}},\ and\ \bibinfo
  {author} {\bibfnamefont {P.}~\bibnamefont {Bambhaniya}},\ }\href
  {https://doi.org/10.1142/S0218271819300246} {\bibfield  {journal} {\bibinfo
  {journal} {Int. J. Mod. Phys. D}\ }\textbf {\bibinfo {volume} {28}},\
  \bibinfo {pages} {1930024} (\bibinfo {year} {2019})},\ \Eprint
  {https://arxiv.org/abs/2101.06001} {arXiv:2101.06001 [gr-qc]} \BibitemShut
  {NoStop}%
\bibitem [{\citenamefont {Bambhaniya}\ \emph {et~al.}(2019)\citenamefont
  {Bambhaniya}, \citenamefont {Joshi}, \citenamefont {Dey},\ and\ \citenamefont
  {Joshi}}]{Bambhaniya:2019pbr}%
  \BibitemOpen
  \bibfield  {author} {\bibinfo {author} {\bibfnamefont {P.}~\bibnamefont
  {Bambhaniya}}, \bibinfo {author} {\bibfnamefont {A.~B.}\ \bibnamefont
  {Joshi}}, \bibinfo {author} {\bibfnamefont {D.}~\bibnamefont {Dey}},\ and\
  \bibinfo {author} {\bibfnamefont {P.~S.}\ \bibnamefont {Joshi}},\ }\href
  {https://doi.org/10.1103/PhysRevD.100.124020} {\bibfield  {journal} {\bibinfo
   {journal} {Phys. Rev. D}\ }\textbf {\bibinfo {volume} {100}},\ \bibinfo
  {pages} {124020} (\bibinfo {year} {2019})},\ \Eprint
  {https://arxiv.org/abs/1908.07171} {arXiv:1908.07171 [gr-qc]} \BibitemShut
  {NoStop}%
\bibitem [{\citenamefont {Joshi}\ \emph {et~al.}(2019)\citenamefont {Joshi},
  \citenamefont {Bambhaniya}, \citenamefont {Dey},\ and\ \citenamefont
  {Joshi}}]{Joshi:2019rdo}%
  \BibitemOpen
  \bibfield  {author} {\bibinfo {author} {\bibfnamefont {A.~B.}\ \bibnamefont
  {Joshi}}, \bibinfo {author} {\bibfnamefont {P.}~\bibnamefont {Bambhaniya}},
  \bibinfo {author} {\bibfnamefont {D.}~\bibnamefont {Dey}},\ and\ \bibinfo
  {author} {\bibfnamefont {P.~S.}\ \bibnamefont {Joshi}},\ }\href@noop {} {\
  (\bibinfo {year} {2019})},\ \Eprint {https://arxiv.org/abs/1909.08873}
  {arXiv:1909.08873 [gr-qc]} \BibitemShut {NoStop}%
\bibitem [{\citenamefont {Dey}\ \emph {et~al.}(2020)\citenamefont {Dey},
  \citenamefont {Shaikh},\ and\ \citenamefont {Joshi}}]{Dey:2020haf}%
  \BibitemOpen
  \bibfield  {author} {\bibinfo {author} {\bibfnamefont {D.}~\bibnamefont
  {Dey}}, \bibinfo {author} {\bibfnamefont {R.}~\bibnamefont {Shaikh}},\ and\
  \bibinfo {author} {\bibfnamefont {P.~S.}\ \bibnamefont {Joshi}},\ }\href
  {https://doi.org/10.1103/PhysRevD.102.044042} {\bibfield  {journal} {\bibinfo
   {journal} {Phys. Rev. D}\ }\textbf {\bibinfo {volume} {102}},\ \bibinfo
  {pages} {044042} (\bibinfo {year} {2020})},\ \Eprint
  {https://arxiv.org/abs/2003.06810} {arXiv:2003.06810 [gr-qc]} \BibitemShut
  {NoStop}%
\bibitem [{\citenamefont {Solanki}\ \emph {et~al.}(2021)\citenamefont
  {Solanki}, \citenamefont {Bambhaniya}, \citenamefont {Dey}, \citenamefont
  {Joshi},\ and\ \citenamefont {Pathak}}]{Solanki:2021mkt}%
  \BibitemOpen
  \bibfield  {author} {\bibinfo {author} {\bibfnamefont {D.~N.}\ \bibnamefont
  {Solanki}}, \bibinfo {author} {\bibfnamefont {P.}~\bibnamefont {Bambhaniya}},
  \bibinfo {author} {\bibfnamefont {D.}~\bibnamefont {Dey}}, \bibinfo {author}
  {\bibfnamefont {P.~S.}\ \bibnamefont {Joshi}},\ and\ \bibinfo {author}
  {\bibfnamefont {K.~N.}\ \bibnamefont {Pathak}},\ }\href@noop {} {\  (\bibinfo
  {year} {2021})},\ \Eprint {https://arxiv.org/abs/2109.14937}
  {arXiv:2109.14937 [gr-qc]} \BibitemShut {NoStop}%
\bibitem [{\citenamefont {Abuter}\ \emph {et~al.}(2017)\citenamefont {Abuter},
  \citenamefont {Accardo}, \citenamefont {Amorim}, \citenamefont {Anugu},
  \citenamefont {Ávila}, \citenamefont {Azouaoui}, \citenamefont {Benisty},
  \citenamefont {Berger}, \citenamefont {Blind},\ and\ \citenamefont
  {et~al.}}]{Abuter:2017}%
  \BibitemOpen
  \bibfield  {author} {\bibinfo {author} {\bibfnamefont {R.}~\bibnamefont
  {Abuter}}, \bibinfo {author} {\bibfnamefont {M.}~\bibnamefont {Accardo}},
  \bibinfo {author} {\bibfnamefont {A.}~\bibnamefont {Amorim}}, \bibinfo
  {author} {\bibfnamefont {N.}~\bibnamefont {Anugu}}, \bibinfo {author}
  {\bibfnamefont {G.}~\bibnamefont {Ávila}}, \bibinfo {author} {\bibfnamefont
  {N.}~\bibnamefont {Azouaoui}}, \bibinfo {author} {\bibfnamefont
  {M.}~\bibnamefont {Benisty}}, \bibinfo {author} {\bibfnamefont {J.~P.}\
  \bibnamefont {Berger}}, \bibinfo {author} {\bibfnamefont {N.}~\bibnamefont
  {Blind}},\ and\ \bibinfo {author} {\bibnamefont {et~al.}},\ }\href
  {https://doi.org/10.1051/0004-6361/201730838} {\bibfield  {journal} {\bibinfo
   {journal} {Astronomy \& Astrophysics}\ }\textbf {\bibinfo {volume} {602}},\
  \bibinfo {pages} {A94} (\bibinfo {year} {2017})}\BibitemShut {NoStop}%
\bibitem [{\citenamefont {Abuter}\ and\ \citenamefont
  {et~al.}(2020)}]{GRAVITY:2020gka}%
  \BibitemOpen
  \bibfield  {author} {\bibinfo {author} {\bibfnamefont {R.}~\bibnamefont
  {Abuter}}\ and\ \bibinfo {author} {\bibnamefont {et~al.}} (\bibinfo
  {collaboration} {GRAVITY}),\ }\href
  {https://doi.org/10.1051/0004-6361/202037813} {\bibfield  {journal} {\bibinfo
   {journal} {Astron. Astrophys.}\ }\textbf {\bibinfo {volume} {636}},\
  \bibinfo {pages} {L5} (\bibinfo {year} {2020})},\ \Eprint
  {https://arxiv.org/abs/2004.07187} {arXiv:2004.07187 [astro-ph.GA]}
  \BibitemShut {NoStop}%
\bibitem [{\citenamefont {Abuter}\ \emph {et~al.}(2019)\citenamefont {Abuter},
  \citenamefont {Amorim}, \citenamefont {Bauböck}, \citenamefont {Berger},
  \citenamefont {Bonnet}, \citenamefont {Brandner}, \citenamefont {Clénet},
  \citenamefont {Coudé~du Foresto}, \citenamefont {de~Zeeuw},\ and\
  \citenamefont {et~al.}}]{Abuter:2019}%
  \BibitemOpen
  \bibfield  {author} {\bibinfo {author} {\bibfnamefont {R.}~\bibnamefont
  {Abuter}}, \bibinfo {author} {\bibfnamefont {A.}~\bibnamefont {Amorim}},
  \bibinfo {author} {\bibfnamefont {M.}~\bibnamefont {Bauböck}}, \bibinfo
  {author} {\bibfnamefont {J.~P.}\ \bibnamefont {Berger}}, \bibinfo {author}
  {\bibfnamefont {H.}~\bibnamefont {Bonnet}}, \bibinfo {author} {\bibfnamefont
  {W.}~\bibnamefont {Brandner}}, \bibinfo {author} {\bibfnamefont
  {Y.}~\bibnamefont {Clénet}}, \bibinfo {author} {\bibfnamefont
  {V.}~\bibnamefont {Coudé~du Foresto}}, \bibinfo {author} {\bibfnamefont
  {P.~T.}\ \bibnamefont {de~Zeeuw}},\ and\ \bibinfo {author} {\bibnamefont
  {et~al.}},\ }\href {https://doi.org/10.1051/0004-6361/201935656} {\bibfield
  {journal} {\bibinfo  {journal} {Astronomy \& Astrophysics}\ }\textbf
  {\bibinfo {volume} {625}},\ \bibinfo {pages} {L10} (\bibinfo {year}
  {2019})}\BibitemShut {NoStop}%
\bibitem [{\citenamefont {Bambhaniya}\ \emph {et~al.}(2021)\citenamefont
  {Bambhaniya}, \citenamefont {Solanki}, \citenamefont {Dey}, \citenamefont
  {Joshi}, \citenamefont {Joshi},\ and\ \citenamefont
  {Patel}}]{Bambhaniya:2020zno}%
  \BibitemOpen
  \bibfield  {author} {\bibinfo {author} {\bibfnamefont {P.}~\bibnamefont
  {Bambhaniya}}, \bibinfo {author} {\bibfnamefont {D.~N.}\ \bibnamefont
  {Solanki}}, \bibinfo {author} {\bibfnamefont {D.}~\bibnamefont {Dey}},
  \bibinfo {author} {\bibfnamefont {A.~B.}\ \bibnamefont {Joshi}}, \bibinfo
  {author} {\bibfnamefont {P.~S.}\ \bibnamefont {Joshi}},\ and\ \bibinfo
  {author} {\bibfnamefont {V.}~\bibnamefont {Patel}},\ }\href
  {https://doi.org/10.1140/epjc/s10052-021-08997-x} {\bibfield  {journal}
  {\bibinfo  {journal} {Eur. Phys. J. C}\ }\textbf {\bibinfo {volume} {81}},\
  \bibinfo {pages} {205} (\bibinfo {year} {2021})},\ \Eprint
  {https://arxiv.org/abs/2007.12086} {arXiv:2007.12086 [gr-qc]} \BibitemShut
  {NoStop}%
\bibitem [{\citenamefont {Virbhadra}\ \emph {et~al.}(1997)\citenamefont
  {Virbhadra}, \citenamefont {Jhingan},\ and\ \citenamefont
  {Joshi}}]{Virbhadra:1995iy}%
  \BibitemOpen
  \bibfield  {author} {\bibinfo {author} {\bibfnamefont {K.~S.}\ \bibnamefont
  {Virbhadra}}, \bibinfo {author} {\bibfnamefont {S.}~\bibnamefont {Jhingan}},\
  and\ \bibinfo {author} {\bibfnamefont {P.~S.}\ \bibnamefont {Joshi}},\ }\href
  {https://doi.org/10.1142/S0218271897000200} {\bibfield  {journal} {\bibinfo
  {journal} {Int. J. Mod. Phys. D}\ }\textbf {\bibinfo {volume} {6}},\ \bibinfo
  {pages} {357} (\bibinfo {year} {1997})},\ \Eprint
  {https://arxiv.org/abs/gr-qc/9512030} {arXiv:gr-qc/9512030} \BibitemShut
  {NoStop}%
\bibitem [{\citenamefont {Zhou}\ \emph {et~al.}(2015)\citenamefont {Zhou},
  \citenamefont {Zhang}, \citenamefont {Chen},\ and\ \citenamefont
  {Wang}}]{Zhou:2014jja}%
  \BibitemOpen
  \bibfield  {author} {\bibinfo {author} {\bibfnamefont {S.}~\bibnamefont
  {Zhou}}, \bibinfo {author} {\bibfnamefont {R.}~\bibnamefont {Zhang}},
  \bibinfo {author} {\bibfnamefont {J.}~\bibnamefont {Chen}},\ and\ \bibinfo
  {author} {\bibfnamefont {Y.}~\bibnamefont {Wang}},\ }\href
  {https://doi.org/10.1007/s10773-015-2526-1} {\bibfield  {journal} {\bibinfo
  {journal} {Int. J. Theor. Phys.}\ }\textbf {\bibinfo {volume} {54}},\
  \bibinfo {pages} {2905} (\bibinfo {year} {2015})},\ \Eprint
  {https://arxiv.org/abs/1408.6041} {arXiv:1408.6041 [gr-qc]} \BibitemShut
  {NoStop}%
\bibitem [{\citenamefont {Darwin}(1959)}]{darwin1959gravity}%
  \BibitemOpen
  \bibfield  {author} {\bibinfo {author} {\bibfnamefont {C.~G.}\ \bibnamefont
  {Darwin}},\ }\href {https://doi.org/10.1098/rspa.1959.0015} {\bibfield
  {journal} {\bibinfo  {journal} {Proceedings of the Royal Society of London.
  Series A. Mathematical and Physical Sciences}\ }\textbf {\bibinfo {volume}
  {249}},\ \bibinfo {pages} {180} (\bibinfo {year} {1959})}\BibitemShut
  {NoStop}%
\bibitem [{\citenamefont {Cutler}\ \emph {et~al.}(1994)\citenamefont {Cutler},
  \citenamefont {Kennefick},\ and\ \citenamefont {Poisson}}]{PhysRevD.50.3816}%
  \BibitemOpen
  \bibfield  {author} {\bibinfo {author} {\bibfnamefont {C.}~\bibnamefont
  {Cutler}}, \bibinfo {author} {\bibfnamefont {D.}~\bibnamefont {Kennefick}},\
  and\ \bibinfo {author} {\bibfnamefont {E.}~\bibnamefont {Poisson}},\ }\href
  {https://doi.org/10.1103/PhysRevD.50.3816} {\bibfield  {journal} {\bibinfo
  {journal} {Phys. Rev. D}\ }\textbf {\bibinfo {volume} {50}},\ \bibinfo
  {pages} {3816} (\bibinfo {year} {1994})}\BibitemShut {NoStop}%
\bibitem [{\citenamefont {Kerner}\ \emph {et~al.}(2001)\citenamefont {Kerner},
  \citenamefont {van Holten},\ and\ \citenamefont {Colistete}}]{Kerner:2001cw}%
  \BibitemOpen
  \bibfield  {author} {\bibinfo {author} {\bibfnamefont {R.}~\bibnamefont
  {Kerner}}, \bibinfo {author} {\bibfnamefont {J.~W.}\ \bibnamefont {van
  Holten}},\ and\ \bibinfo {author} {\bibfnamefont {R.}~\bibnamefont
  {Colistete}, \bibfnamefont {Jr.}},\ }\href
  {https://doi.org/10.1088/0264-9381/18/22/302} {\bibfield  {journal} {\bibinfo
   {journal} {Class. Quant. Grav.}\ }\textbf {\bibinfo {volume} {18}},\
  \bibinfo {pages} {4725} (\bibinfo {year} {2001})},\ \Eprint
  {https://arxiv.org/abs/gr-qc/0102099} {arXiv:gr-qc/0102099} \BibitemShut
  {NoStop}%
\bibitem [{\citenamefont {Philipp}\ \emph {et~al.}(2019)\citenamefont
  {Philipp}, \citenamefont {Puetzfeld},\ and\ \citenamefont
  {L\"ammerzahl}}]{Philipp:2016gyq}%
  \BibitemOpen
  \bibfield  {author} {\bibinfo {author} {\bibfnamefont {D.}~\bibnamefont
  {Philipp}}, \bibinfo {author} {\bibfnamefont {D.}~\bibnamefont {Puetzfeld}},\
  and\ \bibinfo {author} {\bibfnamefont {C.}~\bibnamefont {L\"ammerzahl}},\
  }\href {https://doi.org/10.1007/978-3-030-11500-5_13} {\bibfield  {journal}
  {\bibinfo  {journal} {Fundam. Theor. Phys.}\ }\textbf {\bibinfo {volume}
  {196}},\ \bibinfo {pages} {419} (\bibinfo {year} {2019})},\ \Eprint
  {https://arxiv.org/abs/1604.07173} {arXiv:1604.07173 [gr-qc]} \BibitemShut
  {NoStop}%
\bibitem [{\citenamefont {{Fuchs}}(1990{\natexlab{a}})}]{fuchs1990deviation}%
  \BibitemOpen
  \bibfield  {author} {\bibinfo {author} {\bibfnamefont {H.}~\bibnamefont
  {{Fuchs}}},\ }\href {https://doi.org/10.1002/asna.2113110504} {\bibfield
  {journal} {\bibinfo  {journal} {Astronomische Nachrichten}\ }\textbf
  {\bibinfo {volume} {311}},\ \bibinfo {pages} {271} (\bibinfo {year}
  {1990}{\natexlab{a}})}\BibitemShut {NoStop}%
\bibitem [{\citenamefont
  {{Fuchs}}(1990{\natexlab{b}})}]{fuchs1990paralleltransport}%
  \BibitemOpen
  \bibfield  {author} {\bibinfo {author} {\bibfnamefont {H.}~\bibnamefont
  {{Fuchs}}},\ }\href {https://doi.org/10.1002/asna.2113110405} {\bibfield
  {journal} {\bibinfo  {journal} {Astronomische Nachrichten}\ }\textbf
  {\bibinfo {volume} {311}},\ \bibinfo {pages} {219} (\bibinfo {year}
  {1990}{\natexlab{b}})}\BibitemShut {NoStop}%
\bibitem [{\citenamefont {Kehagias}\ and\ \citenamefont
  {Sfetsos}(2009)}]{Kehagias:2009is}%
  \BibitemOpen
  \bibfield  {author} {\bibinfo {author} {\bibfnamefont {A.}~\bibnamefont
  {Kehagias}}\ and\ \bibinfo {author} {\bibfnamefont {K.}~\bibnamefont
  {Sfetsos}},\ }\href {https://doi.org/10.1016/j.physletb.2009.06.019}
  {\bibfield  {journal} {\bibinfo  {journal} {Phys. Lett. B}\ }\textbf
  {\bibinfo {volume} {678}},\ \bibinfo {pages} {123} (\bibinfo {year}
  {2009})},\ \Eprint {https://arxiv.org/abs/0905.0477} {arXiv:0905.0477
  [hep-th]} \BibitemShut {NoStop}%
\bibitem [{\citenamefont {Ayon-Beato}\ and\ \citenamefont
  {Garcia}(1998)}]{Ayon-Beato:1998hmi}%
  \BibitemOpen
  \bibfield  {author} {\bibinfo {author} {\bibfnamefont {E.}~\bibnamefont
  {Ayon-Beato}}\ and\ \bibinfo {author} {\bibfnamefont {A.}~\bibnamefont
  {Garcia}},\ }\href {https://doi.org/10.1103/PhysRevLett.80.5056} {\bibfield
  {journal} {\bibinfo  {journal} {Phys. Rev. Lett.}\ }\textbf {\bibinfo
  {volume} {80}},\ \bibinfo {pages} {5056} (\bibinfo {year} {1998})},\ \Eprint
  {https://arxiv.org/abs/gr-qc/9911046} {arXiv:gr-qc/9911046} \BibitemShut
  {NoStop}%
\bibitem [{\citenamefont {Nicolini}\ \emph {et~al.}(2006)\citenamefont
  {Nicolini}, \citenamefont {Smailagic},\ and\ \citenamefont
  {Spallucci}}]{Nicolini:2005vd}%
  \BibitemOpen
  \bibfield  {author} {\bibinfo {author} {\bibfnamefont {P.}~\bibnamefont
  {Nicolini}}, \bibinfo {author} {\bibfnamefont {A.}~\bibnamefont
  {Smailagic}},\ and\ \bibinfo {author} {\bibfnamefont {E.}~\bibnamefont
  {Spallucci}},\ }\href {https://doi.org/10.1016/j.physletb.2005.11.004}
  {\bibfield  {journal} {\bibinfo  {journal} {Phys. Lett. B}\ }\textbf
  {\bibinfo {volume} {632}},\ \bibinfo {pages} {547} (\bibinfo {year}
  {2006})},\ \Eprint {https://arxiv.org/abs/gr-qc/0510112}
  {arXiv:gr-qc/0510112} \BibitemShut {NoStop}%
\bibitem [{\citenamefont {Hackmann}\ and\ \citenamefont
  {Lammerzahl}(2008)}]{Hackmann:2008zz}%
  \BibitemOpen
  \bibfield  {author} {\bibinfo {author} {\bibfnamefont {E.}~\bibnamefont
  {Hackmann}}\ and\ \bibinfo {author} {\bibfnamefont {C.}~\bibnamefont
  {Lammerzahl}},\ }\href {https://doi.org/10.1103/PhysRevD.78.024035}
  {\bibfield  {journal} {\bibinfo  {journal} {Phys. Rev. D}\ }\textbf {\bibinfo
  {volume} {78}},\ \bibinfo {pages} {024035} (\bibinfo {year} {2008})},\
  \Eprint {https://arxiv.org/abs/1505.07973} {arXiv:1505.07973 [gr-qc]}
  \BibitemShut {NoStop}%
\bibitem [{\citenamefont {Enolski}\ \emph {et~al.}(2011)\citenamefont
  {Enolski}, \citenamefont {Hackmann}, \citenamefont {Kagramanova},
  \citenamefont {Kunz},\ and\ \citenamefont {Lammerzahl}}]{Enolski:2010if}%
  \BibitemOpen
  \bibfield  {author} {\bibinfo {author} {\bibfnamefont {V.~Z.}\ \bibnamefont
  {Enolski}}, \bibinfo {author} {\bibfnamefont {E.}~\bibnamefont {Hackmann}},
  \bibinfo {author} {\bibfnamefont {V.}~\bibnamefont {Kagramanova}}, \bibinfo
  {author} {\bibfnamefont {J.}~\bibnamefont {Kunz}},\ and\ \bibinfo {author}
  {\bibfnamefont {C.}~\bibnamefont {Lammerzahl}},\ }\href
  {https://doi.org/10.1016/j.geomphys.2011.01.001} {\bibfield  {journal}
  {\bibinfo  {journal} {J. Geom. Phys.}\ }\textbf {\bibinfo {volume} {61}},\
  \bibinfo {pages} {899} (\bibinfo {year} {2011})},\ \Eprint
  {https://arxiv.org/abs/1011.6459} {arXiv:1011.6459 [gr-qc]} \BibitemShut
  {NoStop}%
\end{thebibliography}%

\end{document}